\documentclass[%
 reprint,
 superscriptaddress,
 amsmath,
 amssymb,
 prb,
]{revtex4-2}

\usepackage{graphicx}% Include figure files
\usepackage{dcolumn}% Align table columns on decimal point
\usepackage{bm}% bold math
\usepackage{hyperref}% add hypertext capabilities
\usepackage{siunitx}%for improved typesetting of units
\usepackage{physics}%for improved physics formulas
\usepackage[normalem]{ulem} %Fixes weird linebreak problem in bibliography caused by aps style.

\usepackage{xurl}

\usepackage{qnnmarkup}

\begin{document}

%\preprint{APS/123-QED}

\title{Uncovering Extreme Nonlinear Dynamics in Solids \\ Through Time-Domain Field Analysis}% Force line breaks with \\

\author{P. D. Keathley}
 \thanks{These authors contributed equally.}
 \affiliation{Research Laboratory of Electronics, Massachusetts Institute of Technology, Cambridge, Massachusetts 02139, USA}
\author{S. V. B. Jensen}%
 \thanks{These authors contributed equally.}
 \affiliation{Department of Physics and Astronomy, Aarhus University, DK-8000 Aarhus C, Denmark}
\author{M. Yeung}
 \affiliation{Research Laboratory of Electronics, Massachusetts Institute of Technology, Cambridge, Massachusetts 02139, USA}
\author{M. R. Bionta}
 \affiliation{Research Laboratory of Electronics, Massachusetts Institute of Technology, Cambridge, Massachusetts 02139, USA}
\author{L. B. Madsen}
 \affiliation{Department of Physics and Astronomy, Aarhus University, DK-8000 Aarhus C, Denmark}
\date{\today}% It is always \today, today,

\begin{abstract}
  Time-domain analysis of harmonic fields with sub-cycle resolution is now experimentally viable due to the emergence of sensitive, on-chip techniques for petahertz-scale optical-field sampling. 
  We demonstrate how such a time-domain, field-resolved analysis uncovers the extreme nonlinear electron dynamics responsible for high-harmonic generation within solids.
  Time-dependent density functional theory was used to simulate harmonic generation from a solid-state band-gap system driven by near- to mid-infrared waveforms.
  Particular attention was paid to regimes where both intraband and interband emission mechanisms play a critical role in shaping the nonlinear response.    
  We show that a time-domain analysis of the harmonic radiation fields identifies the interplay between intra- and interband dynamical processes underlying the nonlinear light generation.
  With further analysis, we show that changes to the dominant emission regime can occur after only slight changes to the peak driving intensity and central driving wavelength.
  Time-domain analysis of harmonic fields also reveals, for the first time, the possibility of rapid changes in the dominant emission mechanism within the temporal window of the driving pulse envelope.
  Finally, we examine the experimental viability of performing time-domain analysis of harmonic fields with sub-cycle resolution using realistic parameters.   
\end{abstract}


\maketitle

\section{Introduction}
\label{sec:introduction}

High-harmonic generation (HHG) has proved a fruitful medium for studying extreme nonlinear interactions between intense pulses of light and matter.  Through the study of HHG in atoms and molecules, we have developed a deep understanding of how energy is exchanged between light and electrons on the attosecond timescale~\cite{corkumAttosecondScience2007}. As this work has grown and matured, it has ushered in the rapidly developing fields of attosecond science and technology~\cite{krauszAttosecondPhysics2009}. 

In the past decade researchers have begun exploring HHG from solid-state systems~\cite{ghimireObservationHighorderHarmonic2011, ghimireHighharmonicGenerationSolids2019, goulielmakisHighHarmonicGeneration2022}.
Compared to atomic and molecular systems, solid-state systems open new avenues for fundamental exploration.  
Due to the added role of the solid's bandstructure, and the importance of electron-electron interactions, HHG in solids promises to be an interesting tool for probing attosecond to femtosecond dynamical interactions between light and a variety of materials.
Indeed, HHG in solids has already been used to reconstruct band structure properties of solids \cite{vampaAllOpticalReconstructionCrystal2015,luuExtremeUltravioletHighharmonic2015,uzan-narovlanskyObservationLightdrivenBand2022,uzanAttosecondSpectralSingularities2020a}, measure the Berry curvature \cite{luuMeasurementBerryCurvature2018}, and to track phase transitions in strongly-correlated materials \cite{biontaTrackingUltrafastSolidstate2021}.  
Unfortunately, along with these new avenues for exploration comes additional complexity in experimental analysis.
In particular, both intraband and interband processes result in harmonics with similar non-perturbative intensity scaling laws.  
These similarities make it difficult, and in some cases impossible, to use spectral information alone to disentangle the physical mechanisms underlying HHG in solids. 

Both theoretical and experimental work~\cite{wuHighharmonicGenerationBloch2015, vampaLinkingHighHarmonics2015, vampaAllOpticalReconstructionCrystal2015, hohenleutnerRealtimeObservationInterfering2015, klemkeRoleIntrabandDynamics2020} shows that field-sensitive information is critical to understanding HHG in solids as the harmonic fields carry unique signatures of the underlying emission mechanisms.
While prior work has focused on field-sensitive information gained through phase- and polarization-resolved measurements~\cite{vampaLinkingHighHarmonics2015, vampaAllOpticalReconstructionCrystal2015, hohenleutnerRealtimeObservationInterfering2015, klemkeRoleIntrabandDynamics2020, kobayashiPolarizationFlippingEvenOrder2021},
characterization of the harmonic field waveforms in the time-domain would provide new insights into the underlying dynamics of the generation process. 
Motivated by advances in experimental methods for petahertz-scale optical-field sampling~\cite{herbstRecentAdvancesPetahertz2022a, sederbergAttosecondOptoelectronicField2020, ziminPetahertzscaleNonlinearPhotoconductive2021,ossianderSpeedLimitOptoelectronics2022,liuSingleshotMeasurementFewcycle2022,forgAttosecondNanoscaleNearfield2016,goulielmakisDirectMeasurementLight2004,biontaOnchipSamplingOptical2021}, here we investigate: (1) the physics that would be revealed through a time-domain analysis of the high-harmonic fields generated in solids; and (2) the viability of petahertz-scale field sampling techniques for experimentally measuring harmonic fields directly in the time-domain assuming realistic experimental parameters.   

In Sec.~\ref{sec:tddft-sim-and-analysis} we examine how the interplay between the intra- and interband emission channels manifests in time-domain field signatures.
Using time-dependent density functional theory (TDDFT), we show that the time-domain structure of the generated harmonic fields naturally reveals the dominant HHG emission mechanisms and their corresponding electron dynamics.
We demonstrate how Fourier analysis can be used to break apart and study the temporal structure of both intraband and interband processes in further detail.  
Our analysis reveals that only moderate changes in peak intensity and central driving wavelength can alter the dominant emission mechanism resulting in dramatic changes to the temporal structure of the harmonic fields.  
This dramatic change in the temporal structure of the harmonic fields stands in stark contrast to the minimal changes observed in the HHG spectra.  

Our time-domain field analysis in Sec.~\ref{sec:tddft-sim-and-analysis} also shows that, unlike HHG from gasses, with HHG from solids it cannot be assumed, even for driving pulses containing tens of cycles, that the emitted fields are semi-periodic in time.  
In particular, we observe that under certain conditions the dominant emission mechanism can suddenly switch from intra- to interband over a sub-cycle region of time within the pulse envelope of the driving waveform.  
This complex emission process and resulting lack of periodic structure in the generated harmonic fields means that phase-resolved techniques requiring a certain level of periodicity are in general inadequate for the study of HHG from solids (e.g. techniques similar to reconstruction of attosecond beating by interference of two-photon transitions (RABBITT) that track the interference phase from harmonic to harmonic~\cite{paulObservationTrainAttosecond2001, vampaLinkingHighHarmonics2015}).
We find that measurements having both sub-cycle time resolution and broad spectral coverage are required for a general understanding of solid-state HHG.

%%%%%%%%%%%%%%%%%%%%%%%%%%%%%%%%
%% TDDFT: Compiled Spectra
%%%%%%%%%%%%%%%%%%%%%%%%%%%%%%%% 
\begin{figure}
\includegraphics[width=\columnwidth]{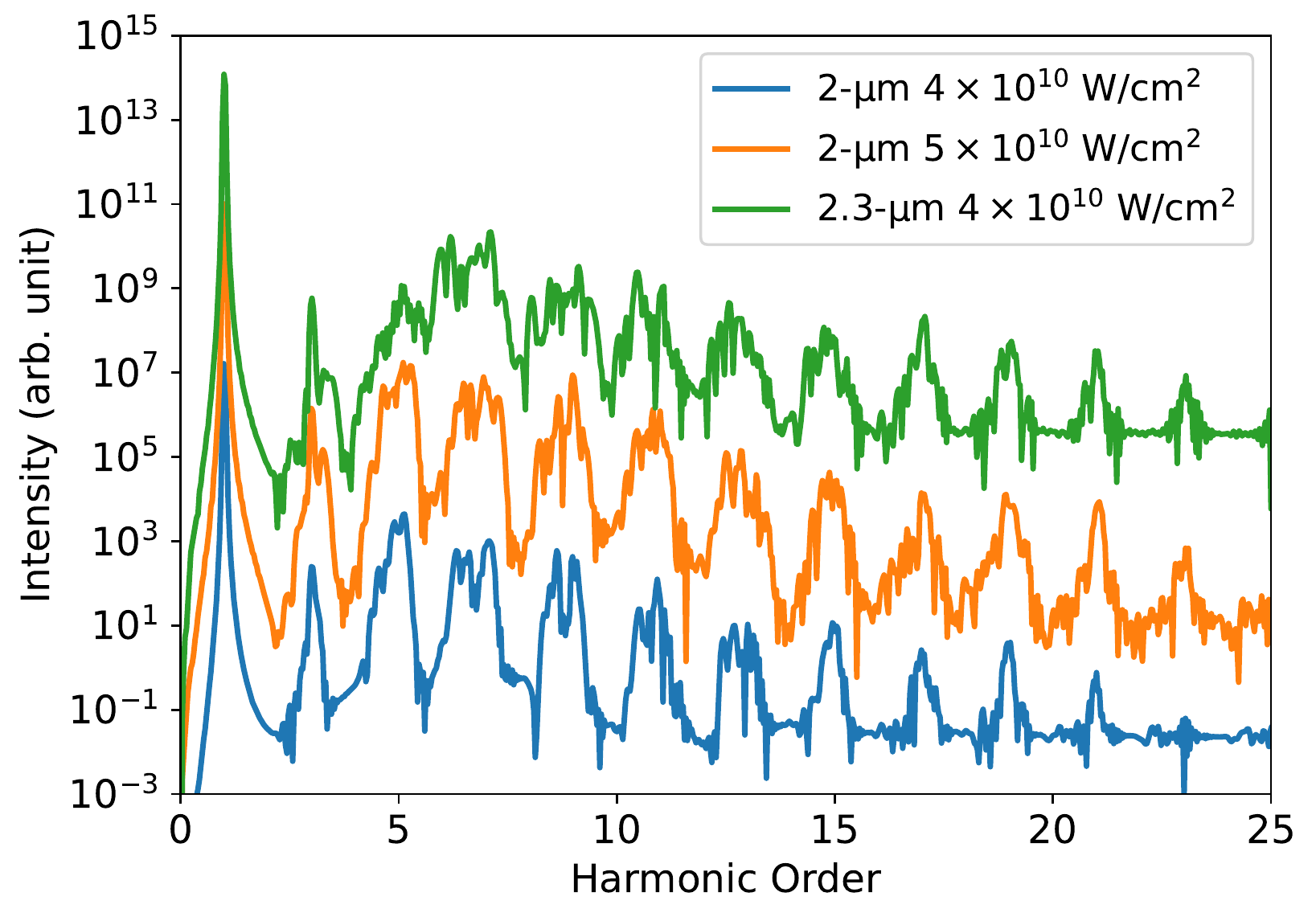}
\caption{\label{fig:compiled-specta} Normalized high harmonic spectra for each condition simulated: driving wavelength of \SI{2}{\micro\meter} with a peak intensity of $4\times10^{10}$~\si{\watt\per\centi\meter^2} (bottom, blue curve); driving wavelength of \SI{2}{\micro\meter} with a peak intensity of $5\times10^{10}$~\si{\watt\per\centi\meter^2} (central, orange curve); and driving wavelength of \SI{2.3}{\micro\meter} with a peak intensity of $4\times10^{10}$~\si{\watt\per\centi\meter^2} (top, green curve).  Each curve is normalized differently for visual clarity.  For each case the same system was used as a generation medium, having a bandgap of roughly $3.3$~\si{\electronvolt}. 
}
\end{figure}

Our findings in Sec.~\ref{sec:tddft-sim-and-analysis} highlight the ripe opportunity for applying recent advances in petahertz-scale field sampling\cite{herbstRecentAdvancesPetahertz2022a} to the characterization and study of HHG in solids.   
In Sec.~\ref{sec:sampling-harmonic-fields} we show that emerging optical field sampling techniques, in particular those that leverage field-driven tunneling ionization~\cite{choTemporalCharacterizationFemtosecond2019,parkDirectSamplingLight2018} and nanostructures for field enhancement~\cite{biontaOnchipSamplingOptical2021,blochlSpatiotemporalSamplingNearpetahertz2022}, place such measurements within reach.
The \textit{ex-situ} nature of these field-sampling approaches makes them ideal for examining the origins of extreme nonlinearities in solids across a variety of material platforms.
Beyond resolving fundamental scientific questions, experimental characterization of harmonic fields will also be vital to the eventual use of HHG from solids in applications designed to leverage the attosecond-scale temporal structure of the harmonic field waveforms in time.      

\section{HHG Simulations and Analysis}
\label{sec:tddft-sim-and-analysis}

Our simulations of HHG in band gap materials are based on time-dependent density functional theory (TDDFT) \cite{rungeDensityFunctionalTheoryTimeDependent1984}. 
It allows for a self-consistent \textit{ab initio} investigation into the dynamics of a laser-driven many-body system. 
It succeeds in accounting for dynamical electron-electron interactions \cite{jensenEdgestateBulklikeLaserinduced2021,tancogne-dejeanParameterfreeHybridlikeFunctional2020} and is suitable for describing experimental features \cite{tancogne-dejeanEllipticityDependenceHighharmonic2017}. Furthermore, TDDFT has been vital to study the mechanisms of HHG in solids \cite{tancogne-dejeanImpactElectronicBand2017} especially as the real-space description allows for introducing dopants \cite{yuEnhancedHighorderHarmonic2019}, topology \cite{bauerHighHarmonicGenerationSolids2018} or finite-system behavior \cite{hansenFinitesystemEffectsHighorder2018,jensenEdgestateinducedCorrelationEffects2021}. For technical details on the TDDFT simulations see Appendix \ref{app:1}. 

We investigate characteristic features of the HHG mechanism using a generic model of a band gap material with a band gap energy of roughly $3.3$~\si{\electronvolt}. We selected two driving wavelengths for comparison: \SI{2}{\micro\meter} ($\approx 0.62$~\si{\electronvolt}) and \SI{2.3}{\micro\meter} ($\approx 0.54$~\si{\electronvolt}) and apply a pulse with a vector potential $A_\text{D}(t)$ having a $30$-cycle $\sin^2$ envelope function. Here the subscript D denotes the driving pulse. For each wavelength a range of intensities were simulated.  We focus on intermediate intensity levels on the order of $10^{10}$~\si{\watt\per\centi\meter^2}. We expect equivalent dynamics for a wide range of systems if modifying the laser parameters to accommodate the specific material properties. 

We start our analysis by examining the harmonic spectra as shown in Fig.~\ref{fig:compiled-specta}.
While there are differences in the spectra, it is impossible from this spectral data alone to determine the interplay of underlying emission mechanisms.
Both intra- and interband processes result in odd-order harmonic generation with non-perturbative scaling of the harmonic strength with intensity.
As we will show, similar-looking spectra resulting from both mechanisms mask dramatic differences in the temporal field structure.

\begin{figure}
\includegraphics[width=\columnwidth]{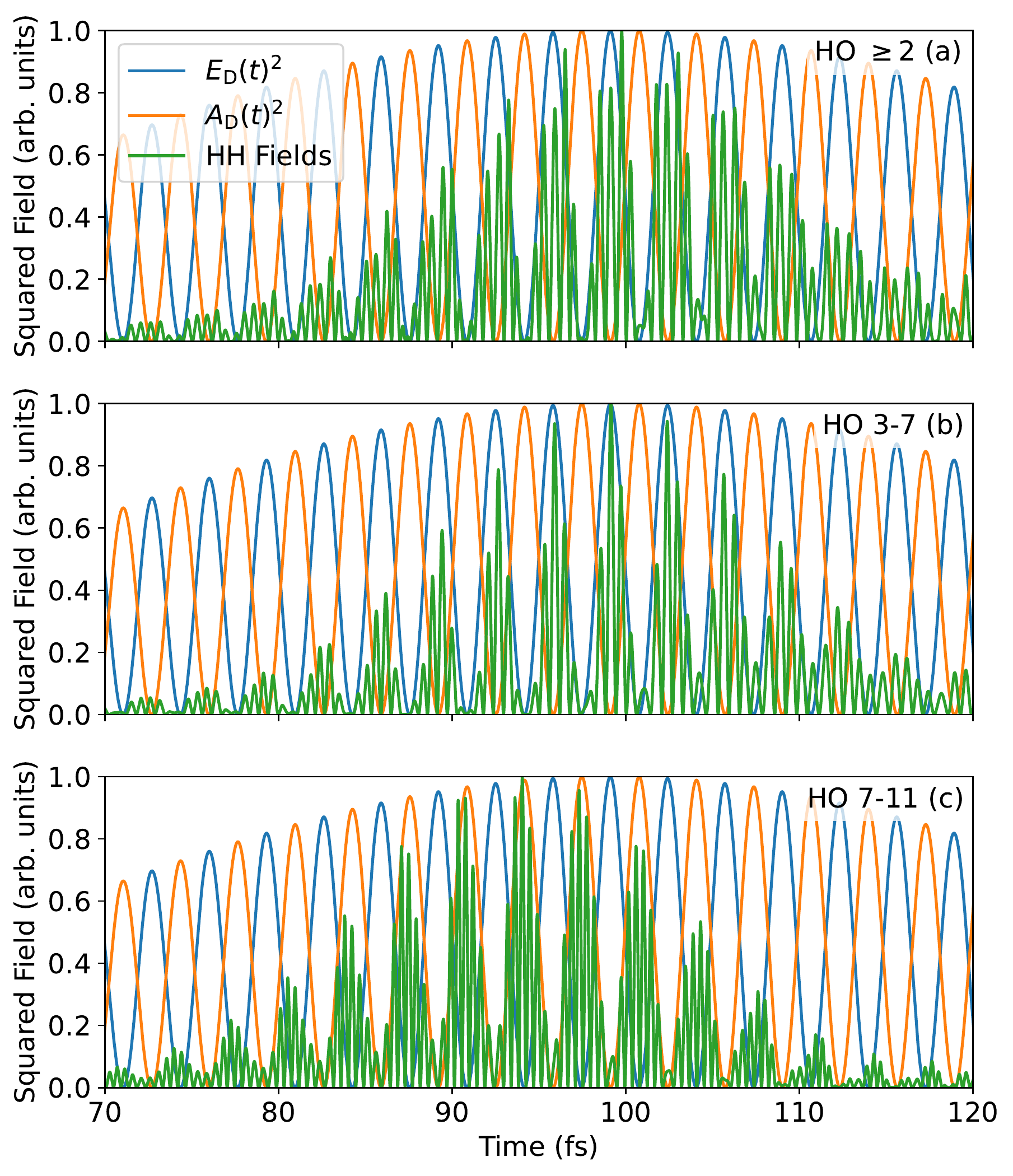}
\caption{\label{fig:time-domain-2-um} Time-domain fields generated by various harmonic contributions for a driving wavelength of \SI{2}{\micro\meter} and peak intensity of $4\times10^{10}$~\si{\watt\per\centi\meter^2}.  (a) All HO $\geq 2$,  (b) HO from 3-7,  and (c) HO from 7-11.}
\end{figure}

For our time-domain analysis, we start by examining the nonlinear system response from the 2-\si{\micro\meter} driver having a peak intensity of $4\times10^{10}$~\si{\watt\per\centi\meter^2} as shown in Fig.~\ref{fig:time-domain-2-um}.
The squared electric field from harmonic orders (HO) $\geq 2$ are shown in green. 
These fields were calculated by applying a high-pass filter to remove the fundamental response oscillating at the driving frequency.
It is useful for analysis to plot the square of the driving field (blue curve) and vector potential (orange curve) as a guide to the eye.  
As has been noted in prior work, intra- and interband responses can be distinguished by their phase relationship with the driving field and vector potential~\cite{hohenleutnerRealtimeObservationInterfering2015,wuHighharmonicGenerationBloch2015}.
Specifically, the interband response is concentrated under the peaks of the squared vector potential (zeros of the squared field), and the intraband response is concentrated under the peaks of the squared driving field (zeros of the vector potential)~\cite{wuHighharmonicGenerationBloch2015}.
Note that for the total nonlinear response in Fig.~\ref{fig:time-domain-2-um}(a), the concentration of harmonic energy aligns with the peaks of the squared electric field is consistent with a dominance of intraband processes in the overall high harmonic emission response.

Through Fourier analysis, it is possible to peel apart the emission behavior even further by examining the time-domain squared field response of selected harmonic regions.
Such analysis is shown in Figs.~\ref{fig:time-domain-2-um}(b) and (c).
In Fig.~\ref{fig:time-domain-2-um}(b) we look at the fields generated only from HO 3-7, and in Fig.~\ref{fig:time-domain-2-um}(c) we look at the fields only from HO 7-11.
Note that we chose to break apart the response at HO 7 as this is just above the band gap energy of the material ($3.3$~\si{\electronvolt} lies between HO 5 and 6).
Consistent with prior theoretical and experimental analysis~\cite{wuHighharmonicGenerationBloch2015,vampaLinkingHighHarmonics2015,hohenleutnerRealtimeObservationInterfering2015}, harmonics below the bandgap have an intraband character with high-harmonic energy concentrated under the peakes of the squared driving field.
This is in stark contrast to harmonics above the bandgap which have an interband character, with high-harmonic energy concentrated under the peaks of the squared vector potential (zeros of the driving field).

%%%%%%%%%%%%%%%%%%%%%%%%%%%%%%%% 
%% TDDFT: 2-um Low 
%%%%%%%%%%%%%%%%%%%%%%%%%%%%%%%% 

\begin{figure}
\includegraphics[width=\columnwidth]{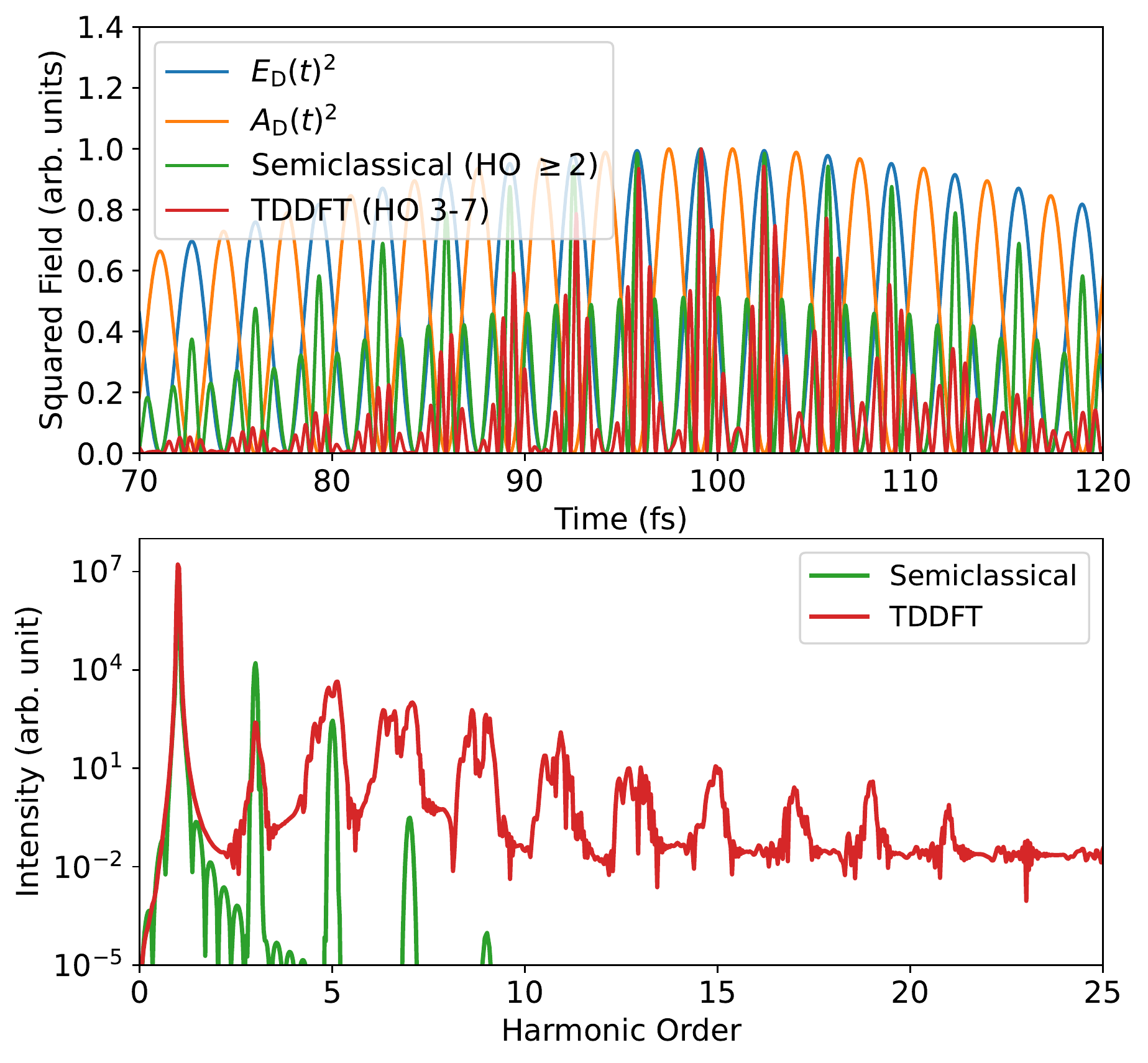}
\caption{\label{fig:tddft-semiclassical-comparison} Comparison between semiclassical and TDDFT model. All driving conditions the same as for Fig.~\ref{fig:time-domain-2-um}. (a) Time-domain fields for all generated HO $\geq 2$ of the semiclassical model (green), and HO 3-7 of the TDDFT model.  Note that just as for the lower HO of the TDDFT model, the semiclassical model predicts HH fields that are concentrated under peaks of the square of the driving electric field.  (b) A comparison of the spectra of the two models.  The semiclassical model is dominated by the lower-order harmonics and drops off rapidly with increasing HO.  On the otherhand, due to the presence of interband processes and more rich accounting of electron-electron interactions, an extended plateau forms in the TDDFT response.  The TDDFT harmonics have rich structuring and a broader bandwidth in comparison with the harmonics predicted by the simple semiclassical response.}
\end{figure}

%%%%%%%%%%%%%%%%%%%%%%%%%%%%%%%% 
%% Description of semi-classical model
%%%%%%%%%%%%%%%%%%%%%%%%%%%%%%%% 
To gain insight into the time-domain behavior of the intraband generation process it is useful to consider the semiclassical model for an intraband electron wavepacket \cite{ashcroftSolidStatePhysics1976,sundaramWavepacketDynamicsSlowly1999}. This model has been applied to understand various features of HHG in solids in the long wavelength regime \cite{ghimireObservationHighorderHarmonic2011,kaneshimaPolarizationResolvedStudyHigh2018,liuHighharmonicGenerationAtomically2017,luuMeasurementBerryCurvature2018, jensenPropagationTimeNondipole2022}. For space and time-inversion symmetric samples, an intraband wavepacket centered at position $x$ and wavevector $k$ is governed by $\dot{k} = - E_\text{D}(t)$ and $ \dot{x} = \pdv*{\varepsilon(k)}{k}$, within the electric-dipole approximation and where we use atomic units. The driving electric field is given by $E_\text{D}(t)=-\pdv*{A_\text{D}(t)}{t}$ and $\varepsilon(k)$ denotes the dispersion, which we extract from the TDDFT calculation.  We consider a single trajectory that is initiated at the $\Gamma$-point with $k(t=0)=0$ and $x(t=0)=0$. Details of the semiclassical modelling are given in Appendix \ref{app:2}.  The generated field $E_{\mathrm{gen},x}(t)$ of the semiclassical intraband electron can be expressed from the current and fulfills the following equation
\begin{equation}
E_{\mathrm{gen},x}(t) \propto \dv{j_x(t)}{t} =   \left(\eval{\dv[2]{\varepsilon(k) }{k}}_{k(t)}\right)  \times E_\mathrm{D}(t), \label{eq:semiclassical}
\end{equation}
where the curvature of the dispersion generally can be expressed from the effective mass tensor $m_\mathrm{eff}^{-1}(k) = \dv*[2]{\varepsilon(k)}{k}$.
An outcome of this is two characteristics of the intraband generation mechanism, (i) the appearance of higher-order harmonics is a result of the non-parabolic curvature of the dispersion (ii) the emission of harmonics in the time-domain is proportional to the driving electric field. This is observed in Fig. \ref{fig:tddft-semiclassical-comparison}, where the emitted field vanishes when the driving electric field vanishes, and has maxima under the driving electric field maxima.

When comparing the semiclassical response to the intraband dominated below bandgap harmonics of the TDDFT calculation in Fig. \ref{fig:tddft-semiclassical-comparison}, one observes that a single trajectory of the semiclassical intraband model is not sufficient to explain the full dynamics of the multi-electron system. If considering the interference of multiple semiclassical trajectories, we expect the spectral features to improve. The phase-relationship with the driving electric field is, however, independent of the initial conditions for the semi-classical trajectories and will thus persist in the interference of multiple trajectories. This characteristic is clearly present in the below bandgap TDDFT dynamics, providing a guideline to decipher signatures of the intraband generation processes in convoluted emission signals.

%%%%%%%%%%%%%%%%%%%%%%%%%%%%%%%% 
%% Comparison: TDDFT and Semiclassical 
%%%%%%%%%%%%%%%%%%%%%%%%%%%%%%%% 

\begin{figure}
\includegraphics[width=\columnwidth]{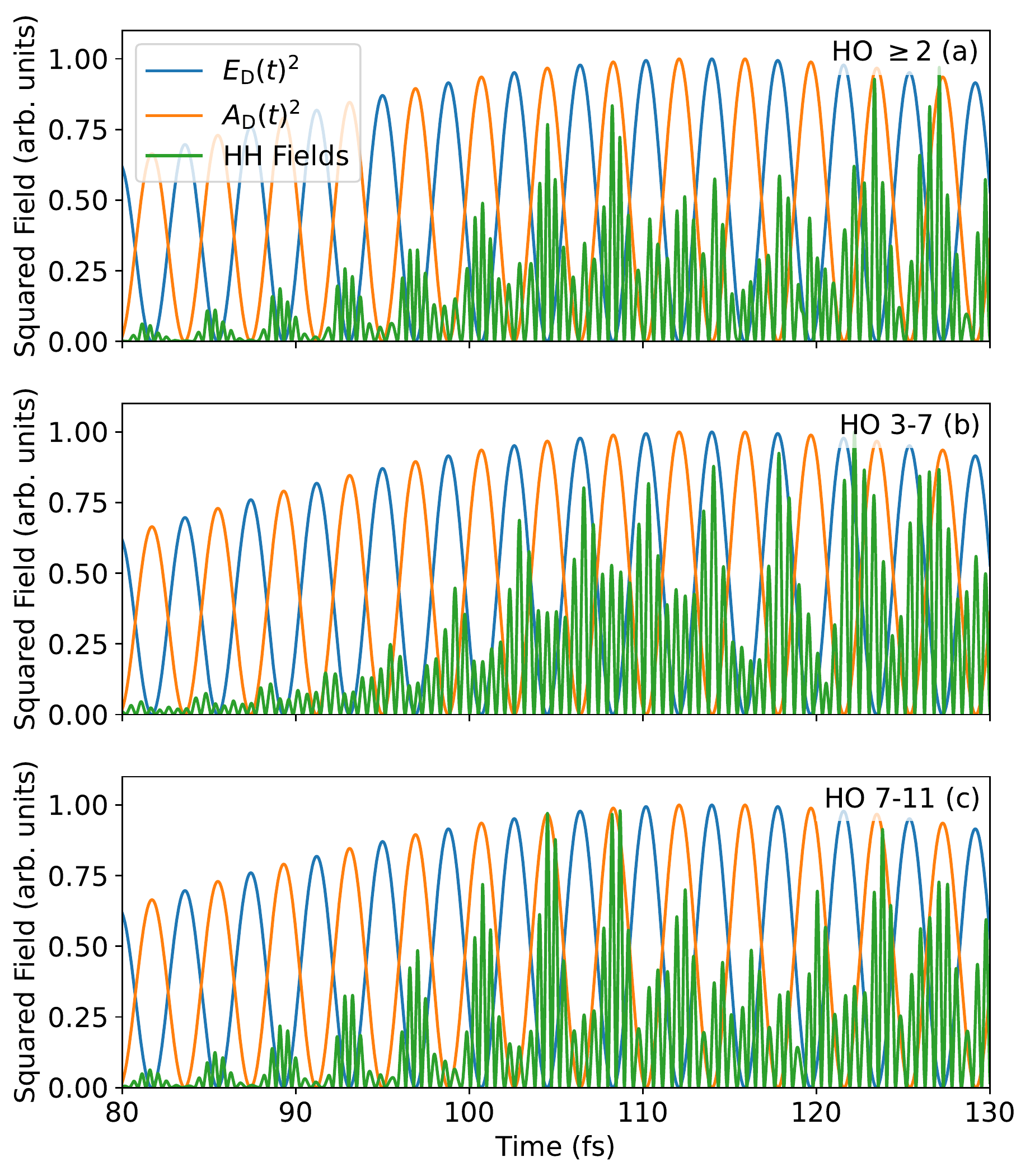}
\caption{\label{fig:time-domain-2p3-um} Time-domain fields generated by various harmonic contributions for a driving wavelength of \SI{2.3}{\micro\meter} and peak intensity $4\times10^{10}$~\si{\watt\per\centi\meter^2}. (a) All HO $\geq 2$,  (b) HO from 3-7,  and (c) HO from 7-11.}
\end{figure}

To study the impact of a change in driving wavelength on the time-domain fields, we repeat the above analysis when keeping the peak intensity fixed at $4\times10^{10}$~\si{\watt\per\centi\meter^2} but with a change of the driving wavelength to \SI{2.3}{\micro\meter} as shown in Fig.~\ref{fig:time-domain-2p3-um}.
Unlike for the 2-\si{\micro\meter} driver, now the emission is dominated by interband-like emission, with the greatest field concentration under the peaks of the squared vector potential.
The analysis of the HO regions generally follows the corresponding response from the 2-\si{\micro\meter} driver, with intraband-like emission dominating emission from HO 3-7, and interband-like emission dominating from HO 7-11.
Referring back to Fig.~\ref{fig:compiled-specta} we note that the overall response is consistent with the shift in strength of the harmonics toward higher HOs for the case of the 2.3-\si{\micro\meter} driver relative to that of the 2-\si{\micro\meter} driver.
Interestingly, the time-domain study also reveals that the harmonic radiation is less concentrated under the peaks of either the squared driving field or squared vector potential as for the case of the 2-\si{\micro\meter} driver. 

%%%%%%%%%%%%%%%%%%%%%%%%%%%%%%%%
%% TDDFT: 2.3-um Low 
%%%%%%%%%%%%%%%%%%%%%%%%%%%%%%%% 

\begin{figure}
\includegraphics[width=\columnwidth]{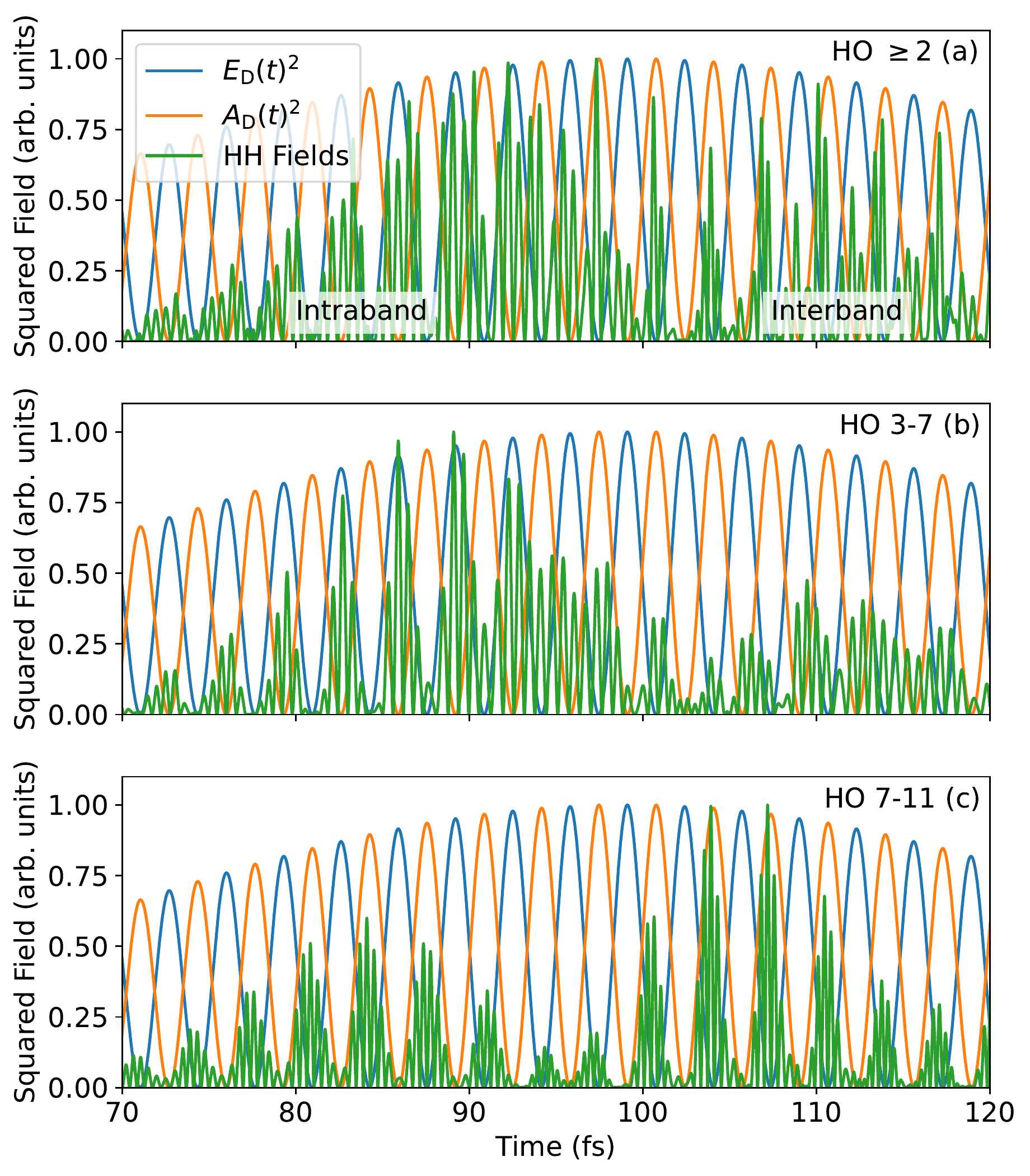}
\caption{\label{fig:time-domain-2-um-high} Time-domain fields generated by various harmonic contributions for a driving wavelength of \SI{2}{\micro\meter} and peak intensity of $5\times10^{10}$~\si{\watt\per\centi\meter^2}. (a) All HO $\geq 2$.  Note the sudden transition from intraband- to interband-type radiation  at $t\approx95$~\si{\femto\second}.  (b) HO from 3-7,  and (c) HO from 7-11.}
\end{figure}

Finally, we study the impact of a change to the peak driving intensity by keeping the driving wavelength fixed at \SI{2}{\micro\meter} while increasing the peak intensity to $5\times10^{10}$~\si{\watt\per\centi\meter^2} as shown in Fig.~\ref{fig:time-domain-2-um-high}.
Interestingly, for this case we observe a dramatic shift in the emission response from intraband-like to interband-like within the time window of the driving pulse envelope.
In Fig.~\ref{fig:time-domain-2-um-high}(a) we see that for times before roughly \SI{96}{\femto\second} the field energy is concentrated mainly under the peaks of the squared driving field, while just after \SI{96}{\femto\second} the fields rapidly concentrate under the peaks of the squared vector potential with a significant reduction in the duration of each half-cycle burst.

Looking at Figs.~\ref{fig:time-domain-2-um-high}(b) and (c), we find on closer inspection that this switch in dominance occurs due to the unique character of the emission response within the time window of the pulse envelope from the two harmonic regions.  At earlier times, HO 3-7, exhibiting intraband-like behavior, dominate the emission response.
The half-cycle emission bursts in this window are concentrated under the peaks of the squared driving field, and experience a sudden drop in intensity near \SI{96}{\femto\second}.
On the other hand, the emission response for HO 7-11 is dominant beyond \SI{96}{\femto\second}.
As with earlier cases, these harmonics have an interband character with the field bursts concentrated under the peaks of the squared vector potential.
They are also shorter in duration than for HO 3-7.

Our findings in this section highlight the need for sub-cycle, field-resolved techniques in order to develop a full understanding of the precise temporal character of the high-harmonic radiation response in solids.

%%%%%%%%%%%%%%%%%%%%%%%%%%%%%%%%
%% TDDFT: 2-um High 
%%%%%%%%%%%%%%%%%%%%%%%%%%%%%%%% 

\section{Sampling HHG Fields}
\label{sec:sampling-harmonic-fields}

In this section we examine the experimental viability of performing time-domain field-analysis of HHG from solids.  
There are two fundamental constraints that have to be satisfied.  
First, adequate temporal resolution of the field measurement process is required.
The needed resolution to capture up the $N^\text{th}$ harmonic would be roughly $T_\text{cyc,D}/(2N)$ where $T_\text{cyc,D}$ is the cycle time of the driving field given by $T_\text{cyc,D} = \lambda_\text{D}/c$, where $\lambda_\text{D}$ is the driving wavelength and $c$ the speed of light. 
For example, assuming a driving wavelength of \SI{2}{\micro\meter}, to capture the time-domain field information up to the 9${}^\text{th}$ harmonic would require a temporal resolution of roughly \SI{370}{\atto\second} or better.   
Second, the technique has to have adequate sensitivity to the signal field.
Given the driving field strengths for the results in Sec.~\ref{sec:tddft-sim-and-analysis} were on the order of \SI{0.5}{\giga\volt\per\meter}, and assuming a generation efficiency on the order of $10^{-6}$, we can estimate that a sensitivity to harmonic field strengths on the order of \si{\mega\volt\per\meter} is needed. 

Several techniques for sampling optical fields with few- to sub-femtosecond resolution have now been demonstrated~\cite{herbstRecentAdvancesPetahertz2022a, sederbergAttosecondOptoelectronicField2020, ziminPetahertzscaleNonlinearPhotoconductive2021,ossianderSpeedLimitOptoelectronics2022,liuSingleshotMeasurementFewcycle2022,keiberElectroopticSamplingNearinfrared2016,forgAttosecondNanoscaleNearfield2016,goulielmakisDirectMeasurementLight2004,biontaOnchipSamplingOptical2021}.
One category of these techniques has shown particular promise for the purpose of sampling HHG from solids: Tunneling Ionization with a Perturbation for the Time-Domain Observation of an Electric Field (TIPTOE)~\cite{choTemporalCharacterizationFemtosecond2019, parkDirectSamplingLight2018}.  
The TIPTOE technique leverages the sub-femtosecond electron tunneling response from gas-phase and solid-state systems driven by strong optical fields ~\cite{parkDirectSamplingLight2018, choTemporalCharacterizationFemtosecond2019, biontaOnchipSamplingOptical2021, blochlSpatiotemporalSamplingNearpetahertz2022,liuSingleshotMeasurementFewcycle2022}.
In the gas phase, TIPTOE has demonstrated the capability of sampling fields with frequencies in excess of \SI{1}{\peta\hertz}~\cite{parkDirectSamplingLight2018} (\textit{i.e.} sub-femtosecond temporal resolution).  
More recently, it has been shown that the sensitivity of TIPTOE can be significantly enhanced using nanostructures~\cite{blochlSpatiotemporalSamplingNearpetahertz2022, biontaOnchipSamplingOptical2021}.  In particular, in Ref.~\cite{biontaOnchipSamplingOptical2021} sensitivity down to \SI{600}{\kilo\volt\per\meter} was demonstrated in the near-infrared using gold nanoantennas.
Encouraged by these demonstrations, in the following we numerically explore the feasibility of a TIPTOE measurement based on optical tunneling from a metal for the examination of the HHG fields analyzed in Sec.~\ref{sec:tddft-sim-and-analysis}.  

\begin{figure*}[t!]
\includegraphics[width=\textwidth]{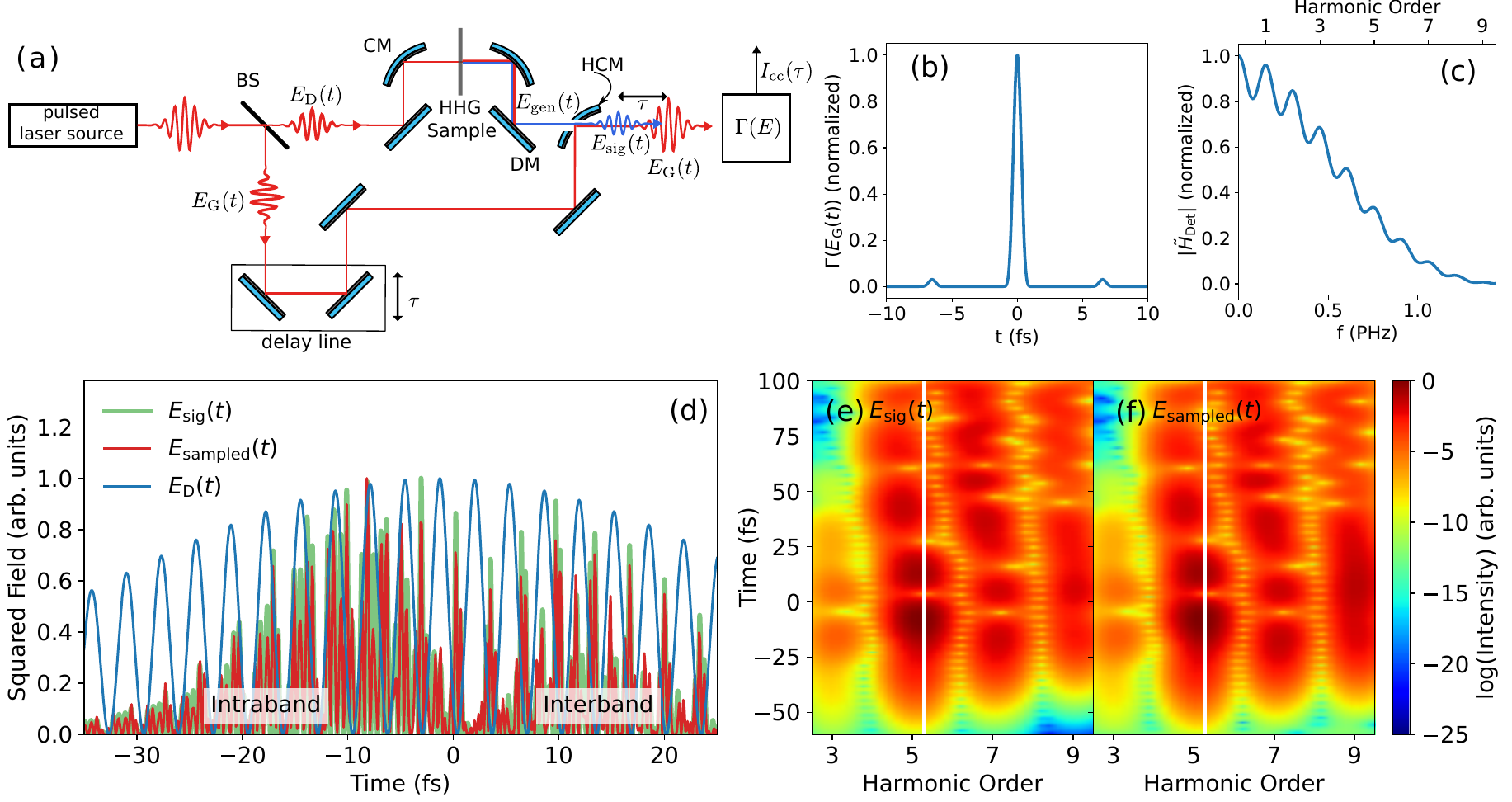}
\caption{\label{fig:sampled-vs-tddft} Simulating harmonic field sampling using solid-state TIPTOE.  (a) Notional schematic of field-sampling experiment using TIPTOE. Abbreviations: BS - beam-splitter; CM - curved mirror; DM - drilled mirror; HCM - holey curved mirror. (b) Calculation of $\Gamma(E_\text{G}(t))$ representing the instantaneous electron tunneling response from the metal surface.  (c) Magnitude of the sampling transfer function $|\tilde{H}_\text{Det}(\omega)|$. (d-f) Comparison of direct TDDFT field output and the result of our simulation of the sampling response. (d) TDDFT output $|E_\text{sig}(t)|^2$ (shaded green) and sampled field response $|E_\text{sampled}(t)|^2$ (solid red) as a function of time.  (e-f) Gabor transform spectrograms of the TDDFT fields (e) and sampled fields (f) showing results up to the $9^{\text{th}}$ harmonic.}
\end{figure*}

We show a notional schematic for the sampling of HHG from a solid-state system using TIPTOE in Fig.~\ref{fig:sampled-vs-tddft}(a).  
An external laser pulse is incident on a beam-splitter (BS) with the transmitted pulse becoming the driving field $E_\text{D}(t)$ for the HHG, and the reflected field becoming the gate field $E_\text{G}(t)$ for driving the TIPTOE measurement.
The generated harmonics are filtered by the dichroic mirror (DM) in order to isolate HO such that measured HO $\geq 2$. 
The reflected harmonic fields become the signal $E_\text{sig}(t)$ for the TIPTOE measurement. In other words, $E_\text{sig}(t)$ is the field resulting from filtering out the fundamental from $E_\text{gen}(t)$.
The delay of $E_\text{G}(t)$ relative to $E_\text{sig}(t)$ is referred to as $\tau$, which is controlled by the lower delay stage.  
For the TIPTOE measurement, one records the oscillations of a cross correlation current $I_\text{cc}(\tau)$ generated via optical-field-driven tunneling from a gas-phase or solid-state system as a function of the delay $\tau$.  
The cross correlation current is approximated as:
\begin{equation}
    I_\text{cc}(\tau) \propto \int_{-\infty}^{\infty}  \left(\left.\dv{\Gamma}{E}\right\vert_{ E_\mathrm{G}(t-\tau)} \right) \times E_\mathrm{sig}(t) \mathrm{d}t,
    \label{eq:cross-corr-current}
\end{equation}
where $\Gamma(E)$ is the tunneling rate as a function of field $E$~\cite{biontaOnchipSamplingOptical2021,choTemporalCharacterizationFemtosecond2019}.  
Here, we model the optical-field-driven tunneling from a metal surface using a Fowler-Nordheim emission rate as described in Refs.~\cite{putnamOpticalfieldcontrolledPhotoemissionPlasmonic2017, biontaOnchipSamplingOptical2021,keathleyVanishingCarrierenvelopephasesensitiveResponse2019}.
We note, however, that other systems, such as gases, semiconductors, or molecules, could be used for generating $I_\text{cc}$.
Our choice was motivated by the ability of metals to achieve significant field enhancements for improved  sensitivity to the signal field~\cite{biontaOnchipSamplingOptical2021,blochlSpatiotemporalSamplingNearpetahertz2022}.  

We start with a gate field $E_\text{G}(t)$ at the metal surface with a work function of \SI{5.1}{\electronvolt}.
The work function chosen is close to that of gold, and representative of most metals, with typical work functions varying between 4 to \SI{6}{\electronvolt}.  
The gate field $E_\text{G}(t)$ was modeled as a Gaussian pulse having a central wavelength of \SI{2}{\micro\meter}, full-width at half-maximum duration of \SI{15}{\femto\second}, and peak field strength of \SI{7.5}{\giga\volt\per\meter}.
The tunneling response from the gate field $\Gamma(E_\text{G}(t))$ is plotted in Fig.~\ref{fig:sampled-vs-tddft}(b).  
The peak gate field strength was chosen to be in line with prior work examining optical tunneling from solids~\cite{putnamOpticalfieldcontrolledPhotoemissionPlasmonic2017,rybkaSubcycleOpticalPhase2016,biontaOnchipSamplingOptical2021,schoetzPerspectivePetahertzElectronics2019} and to provide adequate temporal resolution.
The pulse duration was chosen to ensure a single dominant sub-cycle emission window in time.
We realize that the duration of the gate field is shorter than that of the harmonic driving field provided in Sec.~\ref{sec:tddft-sim-and-analysis}.  
In practice, this could be accommodated via spectral filtering of a short pulse either as part of the transmission response of the beam-splitter, or as a separate element just after the beam-splitter.  

%%%%%%%%%%%%%%%%%%%%%%%%%%%%%%%%
%% Sampled vs. TDDFT Results
%%%%%%%%%%%%%%%%%%%%%%%%%%%%%%%% 

We then take the signal field $E_\text{sig}(t)$ to be that of the harmonic fields as shown in Fig.~\ref{fig:time-domain-2-um-high}, with a peak field strength of \SI{1}{\mega\volt\per\meter}.  We calculate the cross-correlation current $I_\text{cc}(\tau)$ as defined in Eq.~\eqref{eq:cross-corr-current}.  The sampling process imposes its own transfer function $\tilde{H}_\text{Det}(\omega) = \mathcal{F}\Bigl(\left.\dv{\Gamma}{E}\right\vert_{ E_\mathrm{D}(t)}\Bigr) ^*$ which is plotted in Fig.~\ref{fig:sampled-vs-tddft}(c).  
To extract the sampled signal field information from $I_\text{cc}(\tau)$ we then calculated it by taking $E_\text{sampled}(t) = \mathcal{F}^{-1} \Big \lbrace \mathcal{F} \Bigl ( I_\text{cc} \Bigr ) / \tilde{H}_\text{Det}(\omega) \Big \rbrace$.  
We note that in practice the electromagnetic response of the emitter structure (for example, that of a nanoantenna or nanostructured surface if they are being used for increased sensitivity) would also have to be accounted for in the calculation and analysis of $I_\text{cc}(\tau)$.
However, such a broad examination of various systems and their electromagnetic responses is beyond the scope of this work.
Here we focus our attention on fundamental limitations arising from the tunneling response itself.

In Figs.~\ref{fig:sampled-vs-tddft}(d-f) we compare the harmonic fields from Fig.~\ref{fig:time-domain-2-um-high} to the sampled fields calculated through our solid-state TIPTOE analysis.  
Note that the sampled fields reconstruct the input fields with a high degree of accuracy. 
In particular, the sampled field output accurately tracks the sudden transition from intra- to interband-like emission near \SI{96}{\femto\second}.  
In Figs.~\ref{fig:sampled-vs-tddft}(e) and (f) we compare the quality of the sampled data in greater detail using Gabor transformations of the TDDFT and sampled field data.
The tunneling response clearly has sufficient temporal resolution to accurately track the harmonic fields with no loss of information up to the 9${}^{\text{th}}$ harmonic.
However, beyond the 9${}^{\text{th}}$ harmonic, $|\tilde{H}_\text{Det}(\omega)|$ drops off suddenly as shown in Fig.~\ref{fig:sampled-vs-tddft}(c), preventing accurate sampling of higher-frequency fields under the conditions simulated.
However, the bandwidth might be extended by various means.
For example, by using a metal having a higher work function to increase the nonlinearity of the tunneling response, or by using a few-cycle gate pulse with a higher central frequency.

\section{Summary and Conclusion}
\label{sec:summary-and-conclusion}

In this work we have demonstrated how analysis of time-domain fields from HHG clearly reveals the relative contributions of intra- and interband emission mechanisms.
We used TDDFT models that account for harmonic fields generated from both mechanisms within a driven linear atomic chain of atoms.  
We exploited the fact that the fields of intraband harmonics concentrate under the peaks of the squared driving field, while interband harmonics concentrate under the peaks of the squared vector potential to study the interplay between these two emission mechanisms as both a function of HO, as well as in time during the evolution of a single driving pulse.

As with prior work~\cite{wuHighharmonicGenerationBloch2015, vampaLinkingHighHarmonics2015}, we found that harmonics above the bandgap were dominated by interband processes while those below the bandgap were dominated by intraband processes but influenced by the coupling of both intra- and interband processes.  
Using Fourier analysis we were able to isolate and study the temporal response of harmonics from both regimes.
A semiclassical analysis of the intraband generation process provided an emission that was temporally confined to the shape of the driving electric field. This characteristic feature of intraband harmonics was compared to the many-electron TDDFT harmonic emission signal, and found useful as a guideline to decipher harmonics originating from intra- or interband mechanisms.

We also made observations that were uniquely enabled through the analysis of the harmonic fields in time.  
In particular, we observe that complex temporal field structures can arise where intra- and interband dominance can evolve from one temporal region to another within the envelope of the driving pulse.  
Furthermore, this evolution was found to occur over the duration of just a single cycle of the driving field.
Given this observation, we conclude that time-averaged, phase-resolved techniques that assume the harmonics  are well-described as a periodic pulse train having a smooth envelope (e.g. RABBITT and related techniques) are inadequate for a general analysis of the dynamics of HHG in solids.
To properly characterize and analyze rapid shifts in emission dynamics like those observed in Fig.~\ref{fig:time-domain-2-um-high} would necessitate measurements having both sub-cycle time resolution and broad spectral coverage. 

Finally, we explored the feasibility of emerging optical-field sampling methods for the experimental time-domain analysis of high harmonic fields generated in solids.
Given the need for temporal resolution of hundreds of attoseconds and $\sim$\si{\mega\volt\per\meter} field sensitivity, we found the TIPTOE method~\cite{parkDirectSamplingLight2018,choTemporalCharacterizationFemtosecond2019} using a metallic tunneling medium~\cite{blochlSpatiotemporalSamplingNearpetahertz2022,biontaOnchipSamplingOptical2021} to be a compelling choice.  
Our calculations show that under realistic experimental conditions, TIPTOE from metals offers sufficient temporal resolution and spectral coverage to enable time-domain harmonic field analysis.% performed in Sec.~\ref{sec:tddft-sim-and-analysis}.
%However, before experimental realization, various engineering and technical constraints will need to be considered.The electromagnetic response of the tunneling medium (or structure if patterned) will have to be accounted for, and the emitted current and integration times should be adequate for achieving sufficient signal-to-noise ratio.     

Our findings strongly support the pursuit of time-domain field measurements for the analysis of HHG in solids. Methods for sub-cycle field analysis are rapidly advancing~\cite{herbstRecentAdvancesPetahertz2022a}. We are confident that these technologies will become instrumental for uncovering fundamental aspects of  ultrafast strong-field processes in condensed matter directly in the time domain.
%are excited to see what new physics they will uncover about the nature of strong-field processes in condensed matter.

\section{Acknowledgements}
This material is based in part upon work supported by the Air Force Office of Scientific Research under award number FA9550-18-1-0436.
M. Y. Acknowledges support from the National Science Foundation Graduate Research Fellowship Program, Grant No. 1745302.
L. B. M. and S. V. B. J. acknowledges support from the Danish Council for Independent Research (GrantNo.9040-00001B). S. V. B. J. further acknowledges support from the Danish Ministry of Higher Education and Science.

\section{Data and Code Availability}
\label{sec:data-and-code-availability}

The simulation data and code used for analysis and plot generation for this work can be found at \url{https://github.com/qnngroup/manu-uncovering-extreme-nonlinear-dynamics-in-solids-through-time-domain-field-analysis.git}.  

\appendix

\section{TDDFT simulations}\label{app:1}
We represent the electrons by auxiliary noninteracting time-dependent Kohn-Sham (KS) orbitals $\varphi_{\sigma,i} \left(x ,t \right)$ with spin $\sigma =\lbrace\uparrow,\downarrow\rbrace$, which are obtained by imaginary time propagation. When driven by an electromagnetic field, described by the vector potential $A_\text{D}(t)$, the electron dynamics can be captured along the laser polarization direction with a one-dimensional model. Thus, in atomic units, the electron-nuclear interaction is described with a static softened, $\epsilon = 2.25$, Coulomb potential $v_{\mathrm{ion}} \left( x\right) = - \sum_{i=0}^{N-1} Z [\left(x-x_i \right)^2 + \epsilon]^{-1/2}$
formed by a string of $N = 100$ ions of nuclear charge $Z=4$ placed at $x_i  = \left[ i -  \left( N - 1 \right)/2\right] a$ and separated by lattice constant $a= 5.3$.
The KS-orbitals are propagated through the time-dependent KS equation $i \partial_t \varphi_{\sigma,i} \left(x,t \right) = \left\lbrace - \partial^2_x/2 - i A_\text{D}\left(t \right) \partial_x + \tilde{v}_{\mathrm{KS}} \left[n_\sigma\right] \left(x,t \right)  \right\rbrace \varphi_{\sigma,i} \left(x ,t \right)$ using the Crank-Nicolson method with a predictor-corrector step and an absorbing boundary potential \cite{bauerComputationalStrongfieldQuantum2017,kosloffAbsorbingBoundariesWave1986}. The time-dependent KS potential 
$\tilde{v}_{KS}\left[ \left\lbrace n_\sigma \right\rbrace \right] \left( x ,t\right) = v_{\mathrm{ion}} \left(x\right) + v_H \left[ n \right] \left( x ,t \right) + v_{xc}\left[ \left\lbrace n_\sigma \right\rbrace \right] \left( x,t \right)$
contains the Hartree potential $v_H \left[ n \right] \left( x,t \right) = \int dx' n\left( x',t\right)[(x-x')^2 + \epsilon]^{-1/2}$, and the local spin-density exchange-correlation potential $v_{xc}\left[ \left\lbrace n_\sigma \right\rbrace \right] \left( x,t \right) \simeq  - \left[ 6 n_\sigma (x,t) \right/\pi]^{1/3}$. These include the dynamic electron-electron interactions through the density $ n\left(x,t\right) = \sum_{\sigma = \uparrow, \downarrow} n_\sigma \left(x,t\right)$ and spin density $ n_\sigma \left(x,t\right) = \sum_{i=0}^{N_\sigma -1} \abs{\varphi_{\sigma,i} \left( x ,t\right)}^2$. We consider a charge- and spin-neutral system such that the number of electrons with a given spin is $N_{\uparrow, \downarrow} = Z N /2$. The spatial grid contains $21250$ grid points of size $0.1$. Macroscopic propagation effects are not accounted for, as these are suppressed for thin targets \cite{kilenPropagationInducedDephasing2020,yamadaDeterminingOptimumThickness2021}, which can be produced for experiments \cite{liuHighharmonicGenerationAtomically2017}. Similarly, we disregard nondipole effects \cite{jensenPropagationTimeNondipole2022} as these are negligible if describing the generated field as $E_{\mathrm{gen},x}(t) \propto \dv*{j_x(t)}{t}$, with $j_x(t)$ being the current measured along the polarization axis of the driving field. A temporal grid with $95000$ grid points and a step of $0.1$ ensured convergence. 
\section{Semiclassical simulations}\label{app:2}
The intraband generation mechanism is clarified by a numerical simulation of the semiclassical equations. We use the symmetric dispersion in the vicinity of the $\Gamma$-point of the previously obtained TDDFT bandstructure, and expand as a Fourier series $\varepsilon(k) = \left[ 1 + \sum_n c_{n} \cos(n k a) \right]/4a^2$, with coefficients $c_{n} = - 37.4496 \delta_{n,1} +  2.9303  \delta_{n,2} -3.6618 \delta_{n,3} + 0.9685 \delta_{n,4} -1.1009  \delta_{n,5} + 0.5821 \delta_{n,6} -0.4201 \delta_{n,7} + 0.3697 \delta_{n,8}$. We propagate a single-electron wavepacket trajectory of the semiclassical equations, initiated at the $\Gamma$-point, $ k (t=0) = 0$ at $x (t=0) = 0$, as is commonly done in the literature \cite{klemkeRoleIntrabandDynamics2020,luuMeasurementBerryCurvature2018, luuExtremeUltravioletHighharmonic2015}. Explicitly for such material parameters Eq.~\eqref{eq:semiclassical} can be further written as
\begin{equation}
E_{\mathrm{gen},x}(t) \propto  - \left( \sum_n \frac{n^2 c_{n}}{4} \cos[naA_\mathrm{D}(t)] \right) \times   E_\mathrm{D}(t). \label{eq:semiclassical2}
\end{equation}
The first term, arising from the curvature of the dispersion, gives rise to the harmonic contributions in the generated electric field, and is dependent on the initial conditions of the semiclassical wavepacket. The second term is independent of initial conditions, and confines the emitted field to the temporal shape of the driving electric field. 

%\nocite{*}

\bibliography{ref-library}% Produces the bibliography via BibTeX.

\begin{thebibliography}{51}%
\makeatletter
\providecommand \@ifxundefined [1]{%
 \@ifx{#1\undefined}
}%
\providecommand \@ifnum [1]{%
 \ifnum #1\expandafter \@firstoftwo
 \else \expandafter \@secondoftwo
 \fi
}%
\providecommand \@ifx [1]{%
 \ifx #1\expandafter \@firstoftwo
 \else \expandafter \@secondoftwo
 \fi
}%
\providecommand \natexlab [1]{#1}%
\providecommand \enquote  [1]{``#1''}%
\providecommand \bibnamefont  [1]{#1}%
\providecommand \bibfnamefont [1]{#1}%
\providecommand \citenamefont [1]{#1}%
\providecommand \href@noop [0]{\@secondoftwo}%
\providecommand \href [0]{\begingroup \@sanitize@url \@href}%
\providecommand \@href[1]{\@@startlink{#1}\@@href}%
\providecommand \@@href[1]{\endgroup#1\@@endlink}%
\providecommand \@sanitize@url [0]{\catcode `\\12\catcode `\$12\catcode
  `\&12\catcode `\#12\catcode `\^12\catcode `\_12\catcode `\%12\relax}%
\providecommand \@@startlink[1]{}%
\providecommand \@@endlink[0]{}%
\providecommand \url  [0]{\begingroup\@sanitize@url \@url }%
\providecommand \@url [1]{\endgroup\@href {#1}{\urlprefix }}%
\providecommand \urlprefix  [0]{URL }%
\providecommand \Eprint [0]{\href }%
\providecommand \doibase [0]{https://doi.org/}%
\providecommand \selectlanguage [0]{\@gobble}%
\providecommand \bibinfo  [0]{\@secondoftwo}%
\providecommand \bibfield  [0]{\@secondoftwo}%
\providecommand \translation [1]{[#1]}%
\providecommand \BibitemOpen [0]{}%
\providecommand \bibitemStop [0]{}%
\providecommand \bibitemNoStop [0]{.\EOS\space}%
\providecommand \EOS [0]{\spacefactor3000\relax}%
\providecommand \BibitemShut  [1]{\csname bibitem#1\endcsname}%
\let\auto@bib@innerbib\@empty
%</preamble>
\bibitem [{\citenamefont {Corkum}\ and\ \citenamefont
  {Krausz}(2007)}]{corkumAttosecondScience2007}%
  \BibitemOpen
  \bibfield  {author} {\bibinfo {author} {\bibfnamefont {P.~B.}\ \bibnamefont
  {Corkum}}\ and\ \bibinfo {author} {\bibfnamefont {F.}~\bibnamefont
  {Krausz}},\ }\bibfield  {title} {\bibinfo {title} {Attosecond science},\
  }\href {https://doi.org/10.1038/nphys620} {\bibfield  {journal} {\bibinfo
  {journal} {Nat Phys}\ }\textbf {\bibinfo {volume} {3}},\ \bibinfo {pages}
  {381} (\bibinfo {year} {2007})}\BibitemShut {NoStop}%
\bibitem [{\citenamefont {Krausz}\ and\ \citenamefont
  {Ivanov}(2009)}]{krauszAttosecondPhysics2009}%
  \BibitemOpen
  \bibfield  {author} {\bibinfo {author} {\bibfnamefont {F.}~\bibnamefont
  {Krausz}}\ and\ \bibinfo {author} {\bibfnamefont {M.}~\bibnamefont
  {Ivanov}},\ }\bibfield  {title} {\bibinfo {title} {Attosecond physics},\
  }\href {https://doi.org/10.1103/RevModPhys.81.163} {\bibfield  {journal}
  {\bibinfo  {journal} {Reviews of Modern Physics}\ }\textbf {\bibinfo {volume}
  {81}},\ \bibinfo {pages} {163} (\bibinfo {year} {January 00,
  2009})}\BibitemShut {NoStop}%
\bibitem [{\citenamefont {Ghimire}\ \emph {et~al.}(2011)\citenamefont
  {Ghimire}, \citenamefont {DiChiara}, \citenamefont {Sistrunk}, \citenamefont
  {Agostini}, \citenamefont {DiMauro},\ and\ \citenamefont
  {Reis}}]{ghimireObservationHighorderHarmonic2011}%
  \BibitemOpen
  \bibfield  {author} {\bibinfo {author} {\bibfnamefont {S.}~\bibnamefont
  {Ghimire}}, \bibinfo {author} {\bibfnamefont {A.~D.}\ \bibnamefont
  {DiChiara}}, \bibinfo {author} {\bibfnamefont {E.}~\bibnamefont {Sistrunk}},
  \bibinfo {author} {\bibfnamefont {P.}~\bibnamefont {Agostini}}, \bibinfo
  {author} {\bibfnamefont {L.~F.}\ \bibnamefont {DiMauro}},\ and\ \bibinfo
  {author} {\bibfnamefont {D.~A.}\ \bibnamefont {Reis}},\ }\bibfield  {title}
  {\bibinfo {title} {Observation of high-order harmonic generation in a bulk
  crystal},\ }\href {https://doi.org/10.1038/nphys1847} {\bibfield  {journal}
  {\bibinfo  {journal} {Nature Physics}\ }\textbf {\bibinfo {volume} {7}},\
  \bibinfo {pages} {138} (\bibinfo {year} {2011})}\BibitemShut {NoStop}%
\bibitem [{\citenamefont {Ghimire}\ and\ \citenamefont
  {Reis}(2019)}]{ghimireHighharmonicGenerationSolids2019}%
  \BibitemOpen
  \bibfield  {author} {\bibinfo {author} {\bibfnamefont {S.}~\bibnamefont
  {Ghimire}}\ and\ \bibinfo {author} {\bibfnamefont {D.~A.}\ \bibnamefont
  {Reis}},\ }\bibfield  {title} {\bibinfo {title} {High-harmonic generation
  from solids},\ }\href {https://doi.org/10.1038/s41567-018-0315-5} {\bibfield
  {journal} {\bibinfo  {journal} {Nature Physics}\ }\textbf {\bibinfo {volume}
  {15}},\ \bibinfo {pages} {10} (\bibinfo {year} {2019})}\BibitemShut {NoStop}%
\bibitem [{\citenamefont {Goulielmakis}\ and\ \citenamefont
  {Brabec}(2022)}]{goulielmakisHighHarmonicGeneration2022}%
  \BibitemOpen
  \bibfield  {author} {\bibinfo {author} {\bibfnamefont {E.}~\bibnamefont
  {Goulielmakis}}\ and\ \bibinfo {author} {\bibfnamefont {T.}~\bibnamefont
  {Brabec}},\ }\bibfield  {title} {\bibinfo {title} {High harmonic generation
  in condensed matter},\ }\href {https://doi.org/10.1038/s41566-022-00988-y}
  {\bibfield  {journal} {\bibinfo  {journal} {Nature Photonics}\ }\textbf
  {\bibinfo {volume} {16}},\ \bibinfo {pages} {411} (\bibinfo {year}
  {2022})}\BibitemShut {NoStop}%
\bibitem [{\citenamefont {Vampa}\ \emph
  {et~al.}(2015{\natexlab{a}})\citenamefont {Vampa}, \citenamefont {Hammond},
  \citenamefont {Thir{\'e}}, \citenamefont {Schmidt}, \citenamefont
  {L{\'e}gar{\'e}}, \citenamefont {McDonald}, \citenamefont {Brabec},
  \citenamefont {Klug},\ and\ \citenamefont
  {Corkum}}]{vampaAllOpticalReconstructionCrystal2015}%
  \BibitemOpen
  \bibfield  {author} {\bibinfo {author} {\bibfnamefont {G.}~\bibnamefont
  {Vampa}}, \bibinfo {author} {\bibfnamefont {T.~J.}\ \bibnamefont {Hammond}},
  \bibinfo {author} {\bibfnamefont {N.}~\bibnamefont {Thir{\'e}}}, \bibinfo
  {author} {\bibfnamefont {B.~E.}\ \bibnamefont {Schmidt}}, \bibinfo {author}
  {\bibfnamefont {F.}~\bibnamefont {L{\'e}gar{\'e}}}, \bibinfo {author}
  {\bibfnamefont {C.~R.}\ \bibnamefont {McDonald}}, \bibinfo {author}
  {\bibfnamefont {T.}~\bibnamefont {Brabec}}, \bibinfo {author} {\bibfnamefont
  {D.~D.}\ \bibnamefont {Klug}},\ and\ \bibinfo {author} {\bibfnamefont
  {P.~B.}\ \bibnamefont {Corkum}},\ }\bibfield  {title} {\bibinfo {title}
  {All-{{Optical Reconstruction}} of {{Crystal Band Structure}}},\ }\href
  {https://doi.org/10.1103/PhysRevLett.115.193603} {\bibfield  {journal}
  {\bibinfo  {journal} {Physical Review Letters}\ }\textbf {\bibinfo {volume}
  {115}},\ \bibinfo {pages} {193603} (\bibinfo {year}
  {2015}{\natexlab{a}})}\BibitemShut {NoStop}%
\bibitem [{\citenamefont {Luu}\ \emph {et~al.}(2015)\citenamefont {Luu},
  \citenamefont {Garg}, \citenamefont {Kruchinin}, \citenamefont {Moulet},
  \citenamefont {Hassan},\ and\ \citenamefont
  {Goulielmakis}}]{luuExtremeUltravioletHighharmonic2015}%
  \BibitemOpen
  \bibfield  {author} {\bibinfo {author} {\bibfnamefont {T.~T.}\ \bibnamefont
  {Luu}}, \bibinfo {author} {\bibfnamefont {M.}~\bibnamefont {Garg}}, \bibinfo
  {author} {\bibfnamefont {S.~Y.}\ \bibnamefont {Kruchinin}}, \bibinfo {author}
  {\bibfnamefont {A.}~\bibnamefont {Moulet}}, \bibinfo {author} {\bibfnamefont
  {M.~T.}\ \bibnamefont {Hassan}},\ and\ \bibinfo {author} {\bibfnamefont
  {E.}~\bibnamefont {Goulielmakis}},\ }\bibfield  {title} {\bibinfo {title}
  {Extreme ultraviolet high-harmonic spectroscopy of solids},\ }\href
  {https://doi.org/10.1038/nature14456} {\bibfield  {journal} {\bibinfo
  {journal} {Nature}\ }\textbf {\bibinfo {volume} {521}},\ \bibinfo {pages}
  {498} (\bibinfo {year} {2015})}\BibitemShut {NoStop}%
\bibitem [{\citenamefont {{Uzan-Narovlansky}}\ \emph
  {et~al.}(2022)\citenamefont {{Uzan-Narovlansky}}, \citenamefont
  {{Jim{\'e}nez-Gal{\'a}n}}, \citenamefont {Orenstein}, \citenamefont {Silva},
  \citenamefont {{Arusi-Parpar}}, \citenamefont {Shames}, \citenamefont
  {Bruner}, \citenamefont {Yan}, \citenamefont {Smirnova}, \citenamefont
  {Ivanov},\ and\ \citenamefont
  {Dudovich}}]{uzan-narovlanskyObservationLightdrivenBand2022}%
  \BibitemOpen
  \bibfield  {author} {\bibinfo {author} {\bibfnamefont {A.~J.}\ \bibnamefont
  {{Uzan-Narovlansky}}}, \bibinfo {author} {\bibfnamefont {{\'A}.}~\bibnamefont
  {{Jim{\'e}nez-Gal{\'a}n}}}, \bibinfo {author} {\bibfnamefont
  {G.}~\bibnamefont {Orenstein}}, \bibinfo {author} {\bibfnamefont {R.~E.~F.}\
  \bibnamefont {Silva}}, \bibinfo {author} {\bibfnamefont {T.}~\bibnamefont
  {{Arusi-Parpar}}}, \bibinfo {author} {\bibfnamefont {S.}~\bibnamefont
  {Shames}}, \bibinfo {author} {\bibfnamefont {B.~D.}\ \bibnamefont {Bruner}},
  \bibinfo {author} {\bibfnamefont {B.}~\bibnamefont {Yan}}, \bibinfo {author}
  {\bibfnamefont {O.}~\bibnamefont {Smirnova}}, \bibinfo {author}
  {\bibfnamefont {M.}~\bibnamefont {Ivanov}},\ and\ \bibinfo {author}
  {\bibfnamefont {N.}~\bibnamefont {Dudovich}},\ }\bibfield  {title} {\bibinfo
  {title} {Observation of light-driven band structure via multiband
  high-harmonic spectroscopy},\ }\href
  {https://doi.org/10.1038/s41566-022-01010-1} {\bibfield  {journal} {\bibinfo
  {journal} {Nature Photonics}\ }\textbf {\bibinfo {volume} {16}},\ \bibinfo
  {pages} {428} (\bibinfo {year} {2022})}\BibitemShut {NoStop}%
\bibitem [{\citenamefont {Uzan}\ \emph {et~al.}(2020)\citenamefont {Uzan},
  \citenamefont {Orenstein}, \citenamefont {{Jim{\'e}nez-Gal{\'a}n}},
  \citenamefont {McDonald}, \citenamefont {Silva}, \citenamefont {Bruner},
  \citenamefont {Klimkin}, \citenamefont {Blanchet}, \citenamefont
  {{Arusi-Parpar}}, \citenamefont {Kr{\"u}ger}, \citenamefont {Rubtsov},
  \citenamefont {Smirnova}, \citenamefont {Ivanov}, \citenamefont {Yan},
  \citenamefont {Brabec},\ and\ \citenamefont
  {Dudovich}}]{uzanAttosecondSpectralSingularities2020a}%
  \BibitemOpen
  \bibfield  {author} {\bibinfo {author} {\bibfnamefont {A.~J.}\ \bibnamefont
  {Uzan}}, \bibinfo {author} {\bibfnamefont {G.}~\bibnamefont {Orenstein}},
  \bibinfo {author} {\bibfnamefont {{\'A}.}~\bibnamefont
  {{Jim{\'e}nez-Gal{\'a}n}}}, \bibinfo {author} {\bibfnamefont
  {C.}~\bibnamefont {McDonald}}, \bibinfo {author} {\bibfnamefont {R.~E.~F.}\
  \bibnamefont {Silva}}, \bibinfo {author} {\bibfnamefont {B.~D.}\ \bibnamefont
  {Bruner}}, \bibinfo {author} {\bibfnamefont {N.~D.}\ \bibnamefont {Klimkin}},
  \bibinfo {author} {\bibfnamefont {V.}~\bibnamefont {Blanchet}}, \bibinfo
  {author} {\bibfnamefont {T.}~\bibnamefont {{Arusi-Parpar}}}, \bibinfo
  {author} {\bibfnamefont {M.}~\bibnamefont {Kr{\"u}ger}}, \bibinfo {author}
  {\bibfnamefont {A.~N.}\ \bibnamefont {Rubtsov}}, \bibinfo {author}
  {\bibfnamefont {O.}~\bibnamefont {Smirnova}}, \bibinfo {author}
  {\bibfnamefont {M.}~\bibnamefont {Ivanov}}, \bibinfo {author} {\bibfnamefont
  {B.}~\bibnamefont {Yan}}, \bibinfo {author} {\bibfnamefont {T.}~\bibnamefont
  {Brabec}},\ and\ \bibinfo {author} {\bibfnamefont {N.}~\bibnamefont
  {Dudovich}},\ }\bibfield  {title} {\bibinfo {title} {Attosecond spectral
  singularities in solid-state high-harmonic generation},\ }\href
  {https://doi.org/10.1038/s41566-019-0574-4} {\bibfield  {journal} {\bibinfo
  {journal} {Nature Photonics}\ }\textbf {\bibinfo {volume} {14}},\ \bibinfo
  {pages} {183} (\bibinfo {year} {2020})}\BibitemShut {NoStop}%
\bibitem [{\citenamefont {Luu}\ and\ \citenamefont
  {W{\"o}rner}(2018)}]{luuMeasurementBerryCurvature2018}%
  \BibitemOpen
  \bibfield  {author} {\bibinfo {author} {\bibfnamefont {T.~T.}\ \bibnamefont
  {Luu}}\ and\ \bibinfo {author} {\bibfnamefont {H.~J.}\ \bibnamefont
  {W{\"o}rner}},\ }\bibfield  {title} {\bibinfo {title} {Measurement of the
  {{Berry}} curvature of solids using high-harmonic spectroscopy},\ }\href
  {https://doi.org/10.1038/s41467-018-03397-4} {\bibfield  {journal} {\bibinfo
  {journal} {Nature Communications}\ }\textbf {\bibinfo {volume} {9}},\
  \bibinfo {pages} {916} (\bibinfo {year} {2018})}\BibitemShut {NoStop}%
\bibitem [{\citenamefont {Bionta}\ \emph
  {et~al.}(2021{\natexlab{a}})\citenamefont {Bionta}, \citenamefont {Haddad},
  \citenamefont {Leblanc}, \citenamefont {Gruson}, \citenamefont {Lassonde},
  \citenamefont {Ibrahim}, \citenamefont {Chaillou}, \citenamefont {{\'E}mond},
  \citenamefont {Otto}, \citenamefont {{Jim{\'e}nez-Gal{\'a}n}}, \citenamefont
  {Silva}, \citenamefont {Ivanov}, \citenamefont {Siwick}, \citenamefont
  {Chaker},\ and\ \citenamefont
  {L{\'e}gar{\'e}}}]{biontaTrackingUltrafastSolidstate2021}%
  \BibitemOpen
  \bibfield  {author} {\bibinfo {author} {\bibfnamefont {M.~R.}\ \bibnamefont
  {Bionta}}, \bibinfo {author} {\bibfnamefont {E.}~\bibnamefont {Haddad}},
  \bibinfo {author} {\bibfnamefont {A.}~\bibnamefont {Leblanc}}, \bibinfo
  {author} {\bibfnamefont {V.}~\bibnamefont {Gruson}}, \bibinfo {author}
  {\bibfnamefont {P.}~\bibnamefont {Lassonde}}, \bibinfo {author}
  {\bibfnamefont {H.}~\bibnamefont {Ibrahim}}, \bibinfo {author} {\bibfnamefont
  {J.}~\bibnamefont {Chaillou}}, \bibinfo {author} {\bibfnamefont
  {N.}~\bibnamefont {{\'E}mond}}, \bibinfo {author} {\bibfnamefont {M.~R.}\
  \bibnamefont {Otto}}, \bibinfo {author} {\bibfnamefont {{\'A}.}~\bibnamefont
  {{Jim{\'e}nez-Gal{\'a}n}}}, \bibinfo {author} {\bibfnamefont {R.~E.~F.}\
  \bibnamefont {Silva}}, \bibinfo {author} {\bibfnamefont {M.}~\bibnamefont
  {Ivanov}}, \bibinfo {author} {\bibfnamefont {B.~J.}\ \bibnamefont {Siwick}},
  \bibinfo {author} {\bibfnamefont {M.}~\bibnamefont {Chaker}},\ and\ \bibinfo
  {author} {\bibfnamefont {F.}~\bibnamefont {L{\'e}gar{\'e}}},\ }\bibfield
  {title} {\bibinfo {title} {Tracking ultrafast solid-state dynamics using high
  harmonic spectroscopy},\ }\href
  {https://doi.org/10.1103/PhysRevResearch.3.023250} {\bibfield  {journal}
  {\bibinfo  {journal} {Physical Review Research}\ }\textbf {\bibinfo {volume}
  {3}},\ \bibinfo {pages} {023250} (\bibinfo {year}
  {2021}{\natexlab{a}})}\BibitemShut {NoStop}%
\bibitem [{\citenamefont {Wu}\ \emph {et~al.}(2015)\citenamefont {Wu},
  \citenamefont {Ghimire}, \citenamefont {Reis}, \citenamefont {Schafer},\ and\
  \citenamefont {Gaarde}}]{wuHighharmonicGenerationBloch2015}%
  \BibitemOpen
  \bibfield  {author} {\bibinfo {author} {\bibfnamefont {M.}~\bibnamefont
  {Wu}}, \bibinfo {author} {\bibfnamefont {S.}~\bibnamefont {Ghimire}},
  \bibinfo {author} {\bibfnamefont {D.~A.}\ \bibnamefont {Reis}}, \bibinfo
  {author} {\bibfnamefont {K.~J.}\ \bibnamefont {Schafer}},\ and\ \bibinfo
  {author} {\bibfnamefont {M.~B.}\ \bibnamefont {Gaarde}},\ }\bibfield  {title}
  {\bibinfo {title} {High-harmonic generation from {{Bloch}} electrons in
  solids},\ }\href {https://doi.org/10.1103/PhysRevA.91.043839} {\bibfield
  {journal} {\bibinfo  {journal} {Physical Review A}\ }\textbf {\bibinfo
  {volume} {91}},\ \bibinfo {pages} {043839} (\bibinfo {year}
  {2015})}\BibitemShut {NoStop}%
\bibitem [{\citenamefont {Vampa}\ \emph
  {et~al.}(2015{\natexlab{b}})\citenamefont {Vampa}, \citenamefont {Hammond},
  \citenamefont {Thir{\'e}}, \citenamefont {Schmidt}, \citenamefont
  {L{\'e}gar{\'e}}, \citenamefont {McDonald}, \citenamefont {Brabec},\ and\
  \citenamefont {Corkum}}]{vampaLinkingHighHarmonics2015}%
  \BibitemOpen
  \bibfield  {author} {\bibinfo {author} {\bibfnamefont {G.}~\bibnamefont
  {Vampa}}, \bibinfo {author} {\bibfnamefont {T.~J.}\ \bibnamefont {Hammond}},
  \bibinfo {author} {\bibfnamefont {N.}~\bibnamefont {Thir{\'e}}}, \bibinfo
  {author} {\bibfnamefont {B.~E.}\ \bibnamefont {Schmidt}}, \bibinfo {author}
  {\bibfnamefont {F.}~\bibnamefont {L{\'e}gar{\'e}}}, \bibinfo {author}
  {\bibfnamefont {C.~R.}\ \bibnamefont {McDonald}}, \bibinfo {author}
  {\bibfnamefont {T.}~\bibnamefont {Brabec}},\ and\ \bibinfo {author}
  {\bibfnamefont {P.~B.}\ \bibnamefont {Corkum}},\ }\bibfield  {title}
  {\bibinfo {title} {Linking high harmonics from gases and solids},\ }\href
  {https://doi.org/10.1038/nature14517} {\bibfield  {journal} {\bibinfo
  {journal} {Nature}\ }\textbf {\bibinfo {volume} {522}},\ \bibinfo {pages}
  {462} (\bibinfo {year} {2015}{\natexlab{b}})}\BibitemShut {NoStop}%
\bibitem [{\citenamefont {Hohenleutner}\ \emph {et~al.}(2015)\citenamefont
  {Hohenleutner}, \citenamefont {Langer}, \citenamefont {Schubert},
  \citenamefont {Knorr}, \citenamefont {Huttner}, \citenamefont {Koch},
  \citenamefont {Kira},\ and\ \citenamefont
  {Huber}}]{hohenleutnerRealtimeObservationInterfering2015}%
  \BibitemOpen
  \bibfield  {author} {\bibinfo {author} {\bibfnamefont {M.}~\bibnamefont
  {Hohenleutner}}, \bibinfo {author} {\bibfnamefont {F.}~\bibnamefont
  {Langer}}, \bibinfo {author} {\bibfnamefont {O.}~\bibnamefont {Schubert}},
  \bibinfo {author} {\bibfnamefont {M.}~\bibnamefont {Knorr}}, \bibinfo
  {author} {\bibfnamefont {U.}~\bibnamefont {Huttner}}, \bibinfo {author}
  {\bibfnamefont {S.~W.}\ \bibnamefont {Koch}}, \bibinfo {author}
  {\bibfnamefont {M.}~\bibnamefont {Kira}},\ and\ \bibinfo {author}
  {\bibfnamefont {R.}~\bibnamefont {Huber}},\ }\bibfield  {title} {\bibinfo
  {title} {Real-time observation of interfering crystal electrons in
  high-harmonic generation},\ }\href {https://doi.org/10.1038/nature14652}
  {\bibfield  {journal} {\bibinfo  {journal} {Nature}\ }\textbf {\bibinfo
  {volume} {523}},\ \bibinfo {pages} {572} (\bibinfo {year}
  {2015})}\BibitemShut {NoStop}%
\bibitem [{\citenamefont {Klemke}\ \emph {et~al.}(2020)\citenamefont {Klemke},
  \citenamefont {M{\"u}cke}, \citenamefont {Rubio}, \citenamefont
  {K{\"a}rtner},\ and\ \citenamefont
  {{Tancogne-Dejean}}}]{klemkeRoleIntrabandDynamics2020}%
  \BibitemOpen
  \bibfield  {author} {\bibinfo {author} {\bibfnamefont {N.}~\bibnamefont
  {Klemke}}, \bibinfo {author} {\bibfnamefont {O.~D.}\ \bibnamefont
  {M{\"u}cke}}, \bibinfo {author} {\bibfnamefont {A.}~\bibnamefont {Rubio}},
  \bibinfo {author} {\bibfnamefont {F.~X.}\ \bibnamefont {K{\"a}rtner}},\ and\
  \bibinfo {author} {\bibfnamefont {N.}~\bibnamefont {{Tancogne-Dejean}}},\
  }\bibfield  {title} {\bibinfo {title} {Role of intraband dynamics in the
  generation of circularly polarized high harmonics from solids},\ }\href
  {https://doi.org/10.1103/PhysRevB.102.104308} {\bibfield  {journal} {\bibinfo
   {journal} {Physical Review B}\ }\textbf {\bibinfo {volume} {102}},\ \bibinfo
  {pages} {104308} (\bibinfo {year} {2020})}\BibitemShut {NoStop}%
\bibitem [{\citenamefont {Kobayashi}\ \emph {et~al.}(2021)\citenamefont
  {Kobayashi}, \citenamefont {Heide}, \citenamefont {Kelardeh}, \citenamefont
  {Johnson}, \citenamefont {Liu}, \citenamefont {Heinz}, \citenamefont {Reis},\
  and\ \citenamefont {Ghimire}}]{kobayashiPolarizationFlippingEvenOrder2021}%
  \BibitemOpen
  \bibfield  {author} {\bibinfo {author} {\bibfnamefont {Y.}~\bibnamefont
  {Kobayashi}}, \bibinfo {author} {\bibfnamefont {C.}~\bibnamefont {Heide}},
  \bibinfo {author} {\bibfnamefont {H.~K.}\ \bibnamefont {Kelardeh}}, \bibinfo
  {author} {\bibfnamefont {A.}~\bibnamefont {Johnson}}, \bibinfo {author}
  {\bibfnamefont {F.}~\bibnamefont {Liu}}, \bibinfo {author} {\bibfnamefont
  {T.~F.}\ \bibnamefont {Heinz}}, \bibinfo {author} {\bibfnamefont {D.~A.}\
  \bibnamefont {Reis}},\ and\ \bibinfo {author} {\bibfnamefont
  {S.}~\bibnamefont {Ghimire}},\ }\bibfield  {title} {\bibinfo {title}
  {Polarization {{Flipping}} of {{Even-Order Harmonics}} in {{Monolayer
  Transition-Metal Dichalcogenides}}},\ }\bibfield  {journal} {\bibinfo
  {journal} {Ultrafast Science}\ }\textbf {\bibinfo {volume} {2021}},\ \href
  {https://doi.org/10.34133/2021/9820716} {10.34133/2021/9820716} (\bibinfo
  {year} {2021})\BibitemShut {NoStop}%
\bibitem [{\citenamefont {Herbst}\ \emph {et~al.}(2022)\citenamefont {Herbst},
  \citenamefont {Scheffter}, \citenamefont {Bidhendi}, \citenamefont {Kieker},
  \citenamefont {Srivastava},\ and\ \citenamefont
  {Fattahi}}]{herbstRecentAdvancesPetahertz2022a}%
  \BibitemOpen
  \bibfield  {author} {\bibinfo {author} {\bibfnamefont {A.}~\bibnamefont
  {Herbst}}, \bibinfo {author} {\bibfnamefont {K.}~\bibnamefont {Scheffter}},
  \bibinfo {author} {\bibfnamefont {M.~M.}\ \bibnamefont {Bidhendi}}, \bibinfo
  {author} {\bibfnamefont {M.}~\bibnamefont {Kieker}}, \bibinfo {author}
  {\bibfnamefont {A.}~\bibnamefont {Srivastava}},\ and\ \bibinfo {author}
  {\bibfnamefont {H.}~\bibnamefont {Fattahi}},\ }\bibfield  {title} {\bibinfo
  {title} {Recent advances in petahertz electric field sampling},\ }\href
  {https://doi.org/10.1088/1361-6455/ac8032} {\bibfield  {journal} {\bibinfo
  {journal} {Journal of Physics B: Atomic, Molecular and Optical Physics}\
  }\textbf {\bibinfo {volume} {55}},\ \bibinfo {pages} {172001} (\bibinfo
  {year} {2022})}\BibitemShut {NoStop}%
\bibitem [{\citenamefont {Sederberg}\ \emph {et~al.}(2020)\citenamefont
  {Sederberg}, \citenamefont {Zimin}, \citenamefont {Keiber}, \citenamefont
  {Siegrist}, \citenamefont {Wismer}, \citenamefont {Yakovlev}, \citenamefont
  {Floss}, \citenamefont {Lemell}, \citenamefont {Burgd{\"o}rfer},
  \citenamefont {Schultze}, \citenamefont {Krausz},\ and\ \citenamefont
  {Karpowicz}}]{sederbergAttosecondOptoelectronicField2020}%
  \BibitemOpen
  \bibfield  {author} {\bibinfo {author} {\bibfnamefont {S.}~\bibnamefont
  {Sederberg}}, \bibinfo {author} {\bibfnamefont {D.}~\bibnamefont {Zimin}},
  \bibinfo {author} {\bibfnamefont {S.}~\bibnamefont {Keiber}}, \bibinfo
  {author} {\bibfnamefont {F.}~\bibnamefont {Siegrist}}, \bibinfo {author}
  {\bibfnamefont {M.~S.}\ \bibnamefont {Wismer}}, \bibinfo {author}
  {\bibfnamefont {V.~S.}\ \bibnamefont {Yakovlev}}, \bibinfo {author}
  {\bibfnamefont {I.}~\bibnamefont {Floss}}, \bibinfo {author} {\bibfnamefont
  {C.}~\bibnamefont {Lemell}}, \bibinfo {author} {\bibfnamefont
  {J.}~\bibnamefont {Burgd{\"o}rfer}}, \bibinfo {author} {\bibfnamefont
  {M.}~\bibnamefont {Schultze}}, \bibinfo {author} {\bibfnamefont
  {F.}~\bibnamefont {Krausz}},\ and\ \bibinfo {author} {\bibfnamefont
  {N.}~\bibnamefont {Karpowicz}},\ }\bibfield  {title} {\bibinfo {title}
  {Attosecond optoelectronic field measurement in solids},\ }\href
  {https://doi.org/10.1038/s41467-019-14268-x} {\bibfield  {journal} {\bibinfo
  {journal} {Nature Communications}\ }\textbf {\bibinfo {volume} {11}},\
  \bibinfo {pages} {1} (\bibinfo {year} {2020})}\BibitemShut {NoStop}%
\bibitem [{\citenamefont {Zimin}\ \emph {et~al.}(2021)\citenamefont {Zimin},
  \citenamefont {Weidman}, \citenamefont {Sch{\"o}tz}, \citenamefont {Kling},
  \citenamefont {Yakovlev}, \citenamefont {Krausz},\ and\ \citenamefont
  {Karpowicz}}]{ziminPetahertzscaleNonlinearPhotoconductive2021}%
  \BibitemOpen
  \bibfield  {author} {\bibinfo {author} {\bibfnamefont {D.}~\bibnamefont
  {Zimin}}, \bibinfo {author} {\bibfnamefont {M.}~\bibnamefont {Weidman}},
  \bibinfo {author} {\bibfnamefont {J.}~\bibnamefont {Sch{\"o}tz}}, \bibinfo
  {author} {\bibfnamefont {M.~F.}\ \bibnamefont {Kling}}, \bibinfo {author}
  {\bibfnamefont {V.~S.}\ \bibnamefont {Yakovlev}}, \bibinfo {author}
  {\bibfnamefont {F.}~\bibnamefont {Krausz}},\ and\ \bibinfo {author}
  {\bibfnamefont {N.}~\bibnamefont {Karpowicz}},\ }\bibfield  {title} {\bibinfo
  {title} {Petahertz-scale nonlinear photoconductive sampling in air},\ }\href
  {https://doi.org/10.1364/OPTICA.411434} {\bibfield  {journal} {\bibinfo
  {journal} {Optica}\ }\textbf {\bibinfo {volume} {8}},\ \bibinfo {pages} {586}
  (\bibinfo {year} {2021})}\BibitemShut {NoStop}%
\bibitem [{\citenamefont {Ossiander}\ \emph {et~al.}(2022)\citenamefont
  {Ossiander}, \citenamefont {Golyari}, \citenamefont {Scharl}, \citenamefont
  {Lehnert}, \citenamefont {Siegrist}, \citenamefont {B{\"u}rger},
  \citenamefont {Zimin}, \citenamefont {Gessner}, \citenamefont {Weidman},
  \citenamefont {Floss}, \citenamefont {Smejkal}, \citenamefont {Donsa},
  \citenamefont {Lemell}, \citenamefont {Libisch}, \citenamefont {Karpowicz},
  \citenamefont {Burgd{\"o}rfer}, \citenamefont {Krausz},\ and\ \citenamefont
  {Schultze}}]{ossianderSpeedLimitOptoelectronics2022}%
  \BibitemOpen
  \bibfield  {author} {\bibinfo {author} {\bibfnamefont {M.}~\bibnamefont
  {Ossiander}}, \bibinfo {author} {\bibfnamefont {K.}~\bibnamefont {Golyari}},
  \bibinfo {author} {\bibfnamefont {K.}~\bibnamefont {Scharl}}, \bibinfo
  {author} {\bibfnamefont {L.}~\bibnamefont {Lehnert}}, \bibinfo {author}
  {\bibfnamefont {F.}~\bibnamefont {Siegrist}}, \bibinfo {author}
  {\bibfnamefont {J.~P.}\ \bibnamefont {B{\"u}rger}}, \bibinfo {author}
  {\bibfnamefont {D.}~\bibnamefont {Zimin}}, \bibinfo {author} {\bibfnamefont
  {J.~A.}\ \bibnamefont {Gessner}}, \bibinfo {author} {\bibfnamefont
  {M.}~\bibnamefont {Weidman}}, \bibinfo {author} {\bibfnamefont
  {I.}~\bibnamefont {Floss}}, \bibinfo {author} {\bibfnamefont
  {V.}~\bibnamefont {Smejkal}}, \bibinfo {author} {\bibfnamefont
  {S.}~\bibnamefont {Donsa}}, \bibinfo {author} {\bibfnamefont
  {C.}~\bibnamefont {Lemell}}, \bibinfo {author} {\bibfnamefont
  {F.}~\bibnamefont {Libisch}}, \bibinfo {author} {\bibfnamefont
  {N.}~\bibnamefont {Karpowicz}}, \bibinfo {author} {\bibfnamefont
  {J.}~\bibnamefont {Burgd{\"o}rfer}}, \bibinfo {author} {\bibfnamefont
  {F.}~\bibnamefont {Krausz}},\ and\ \bibinfo {author} {\bibfnamefont
  {M.}~\bibnamefont {Schultze}},\ }\bibfield  {title} {\bibinfo {title} {The
  speed limit of optoelectronics},\ }\href
  {https://doi.org/10.1038/s41467-022-29252-1} {\bibfield  {journal} {\bibinfo
  {journal} {Nature Communications}\ }\textbf {\bibinfo {volume} {13}},\
  \bibinfo {pages} {1620} (\bibinfo {year} {2022})}\BibitemShut {NoStop}%
\bibitem [{\citenamefont {Liu}\ \emph {et~al.}(2022)\citenamefont {Liu},
  \citenamefont {Beetar}, \citenamefont {Nesper}, \citenamefont
  {{Gholam-Mirzaei}},\ and\ \citenamefont
  {Chini}}]{liuSingleshotMeasurementFewcycle2022}%
  \BibitemOpen
  \bibfield  {author} {\bibinfo {author} {\bibfnamefont {Y.}~\bibnamefont
  {Liu}}, \bibinfo {author} {\bibfnamefont {J.~E.}\ \bibnamefont {Beetar}},
  \bibinfo {author} {\bibfnamefont {J.}~\bibnamefont {Nesper}}, \bibinfo
  {author} {\bibfnamefont {S.}~\bibnamefont {{Gholam-Mirzaei}}},\ and\ \bibinfo
  {author} {\bibfnamefont {M.}~\bibnamefont {Chini}},\ }\bibfield  {title}
  {\bibinfo {title} {Single-shot measurement of few-cycle optical waveforms on
  a chip},\ }\href {https://doi.org/10.1038/s41566-021-00924-6} {\bibfield
  {journal} {\bibinfo  {journal} {Nature Photonics}\ }\textbf {\bibinfo
  {volume} {16}},\ \bibinfo {pages} {109} (\bibinfo {year} {2022})}\BibitemShut
  {NoStop}%
\bibitem [{\citenamefont {F{\"o}rg}\ \emph {et~al.}(2016)\citenamefont
  {F{\"o}rg}, \citenamefont {Sch{\"o}tz}, \citenamefont {S{\"u}{\ss}mann},
  \citenamefont {F{\"o}rster}, \citenamefont {Kr{\"u}ger}, \citenamefont {Ahn},
  \citenamefont {Okell}, \citenamefont {Wintersperger}, \citenamefont
  {Zherebtsov}, \citenamefont {Guggenmos}, \citenamefont {Pervak},
  \citenamefont {Kessel}, \citenamefont {Trushin}, \citenamefont {Azzeer},
  \citenamefont {Stockman}, \citenamefont {Kim}, \citenamefont {Krausz},
  \citenamefont {Hommelhoff},\ and\ \citenamefont
  {Kling}}]{forgAttosecondNanoscaleNearfield2016}%
  \BibitemOpen
  \bibfield  {author} {\bibinfo {author} {\bibfnamefont {B.}~\bibnamefont
  {F{\"o}rg}}, \bibinfo {author} {\bibfnamefont {J.}~\bibnamefont
  {Sch{\"o}tz}}, \bibinfo {author} {\bibfnamefont {F.}~\bibnamefont
  {S{\"u}{\ss}mann}}, \bibinfo {author} {\bibfnamefont {M.}~\bibnamefont
  {F{\"o}rster}}, \bibinfo {author} {\bibfnamefont {M.}~\bibnamefont
  {Kr{\"u}ger}}, \bibinfo {author} {\bibfnamefont {B.}~\bibnamefont {Ahn}},
  \bibinfo {author} {\bibfnamefont {W.~A.}\ \bibnamefont {Okell}}, \bibinfo
  {author} {\bibfnamefont {K.}~\bibnamefont {Wintersperger}}, \bibinfo {author}
  {\bibfnamefont {S.}~\bibnamefont {Zherebtsov}}, \bibinfo {author}
  {\bibfnamefont {A.}~\bibnamefont {Guggenmos}}, \bibinfo {author}
  {\bibfnamefont {V.}~\bibnamefont {Pervak}}, \bibinfo {author} {\bibfnamefont
  {A.}~\bibnamefont {Kessel}}, \bibinfo {author} {\bibfnamefont {S.~A.}\
  \bibnamefont {Trushin}}, \bibinfo {author} {\bibfnamefont {A.~M.}\
  \bibnamefont {Azzeer}}, \bibinfo {author} {\bibfnamefont {M.~I.}\
  \bibnamefont {Stockman}}, \bibinfo {author} {\bibfnamefont {D.}~\bibnamefont
  {Kim}}, \bibinfo {author} {\bibfnamefont {F.}~\bibnamefont {Krausz}},
  \bibinfo {author} {\bibfnamefont {P.}~\bibnamefont {Hommelhoff}},\ and\
  \bibinfo {author} {\bibfnamefont {M.~F.}\ \bibnamefont {Kling}},\ }\bibfield
  {title} {\bibinfo {title} {Attosecond nanoscale near-field sampling},\ }\href
  {https://doi.org/10.1038/ncomms11717} {\bibfield  {journal} {\bibinfo
  {journal} {Nature Communications}\ }\textbf {\bibinfo {volume} {7}},\
  \bibinfo {pages} {11717} (\bibinfo {year} {2016})}\BibitemShut {NoStop}%
\bibitem [{\citenamefont {Goulielmakis}\ \emph {et~al.}(2004)\citenamefont
  {Goulielmakis}, \citenamefont {Uiberacker}, \citenamefont {Kienberger},
  \citenamefont {Baltuska}, \citenamefont {Yakovlev}, \citenamefont {Scrinzi},
  \citenamefont {Westerwalbesloh}, \citenamefont {Kleineberg}, \citenamefont
  {Heinzmann}, \citenamefont {Drescher},\ and\ \citenamefont
  {Krausz}}]{goulielmakisDirectMeasurementLight2004}%
  \BibitemOpen
  \bibfield  {author} {\bibinfo {author} {\bibfnamefont {E.}~\bibnamefont
  {Goulielmakis}}, \bibinfo {author} {\bibfnamefont {M.}~\bibnamefont
  {Uiberacker}}, \bibinfo {author} {\bibfnamefont {R.}~\bibnamefont
  {Kienberger}}, \bibinfo {author} {\bibfnamefont {A.}~\bibnamefont
  {Baltuska}}, \bibinfo {author} {\bibfnamefont {V.}~\bibnamefont {Yakovlev}},
  \bibinfo {author} {\bibfnamefont {A.}~\bibnamefont {Scrinzi}}, \bibinfo
  {author} {\bibfnamefont {T.}~\bibnamefont {Westerwalbesloh}}, \bibinfo
  {author} {\bibfnamefont {U.}~\bibnamefont {Kleineberg}}, \bibinfo {author}
  {\bibfnamefont {U.}~\bibnamefont {Heinzmann}}, \bibinfo {author}
  {\bibfnamefont {M.}~\bibnamefont {Drescher}},\ and\ \bibinfo {author}
  {\bibfnamefont {F.}~\bibnamefont {Krausz}},\ }\bibfield  {title} {\bibinfo
  {title} {Direct {{Measurement}} of {{Light Waves}}},\ }\href
  {https://doi.org/10.1126/science.1100866} {\bibfield  {journal} {\bibinfo
  {journal} {Science}\ }\textbf {\bibinfo {volume} {305}},\ \bibinfo {pages}
  {1267} (\bibinfo {year} {2004})}\BibitemShut {NoStop}%
\bibitem [{\citenamefont {Bionta}\ \emph
  {et~al.}(2021{\natexlab{b}})\citenamefont {Bionta}, \citenamefont
  {Ritzkowsky}, \citenamefont {Turchetti}, \citenamefont {Yang}, \citenamefont
  {Cattozzo~Mor}, \citenamefont {Putnam}, \citenamefont {K{\"a}rtner},
  \citenamefont {Berggren},\ and\ \citenamefont
  {Keathley}}]{biontaOnchipSamplingOptical2021}%
  \BibitemOpen
  \bibfield  {author} {\bibinfo {author} {\bibfnamefont {M.~R.}\ \bibnamefont
  {Bionta}}, \bibinfo {author} {\bibfnamefont {F.}~\bibnamefont {Ritzkowsky}},
  \bibinfo {author} {\bibfnamefont {M.}~\bibnamefont {Turchetti}}, \bibinfo
  {author} {\bibfnamefont {Y.}~\bibnamefont {Yang}}, \bibinfo {author}
  {\bibfnamefont {D.}~\bibnamefont {Cattozzo~Mor}}, \bibinfo {author}
  {\bibfnamefont {W.~P.}\ \bibnamefont {Putnam}}, \bibinfo {author}
  {\bibfnamefont {F.~X.}\ \bibnamefont {K{\"a}rtner}}, \bibinfo {author}
  {\bibfnamefont {K.~K.}\ \bibnamefont {Berggren}},\ and\ \bibinfo {author}
  {\bibfnamefont {P.~D.}\ \bibnamefont {Keathley}},\ }\bibfield  {title}
  {\bibinfo {title} {On-chip sampling of optical fields with attosecond
  resolution},\ }\href {https://doi.org/10.1038/s41566-021-00792-0} {\bibfield
  {journal} {\bibinfo  {journal} {Nature Photonics}\ }\textbf {\bibinfo
  {volume} {15}},\ \bibinfo {pages} {456} (\bibinfo {year}
  {2021}{\natexlab{b}})}\BibitemShut {NoStop}%
\bibitem [{\citenamefont {Paul}\ \emph {et~al.}(2001)\citenamefont {Paul},
  \citenamefont {Toma}, \citenamefont {Breger}, \citenamefont {Mullot},
  \citenamefont {Auge}, \citenamefont {Balcou}, \citenamefont {Muller},\ and\
  \citenamefont {Agostini}}]{paulObservationTrainAttosecond2001}%
  \BibitemOpen
  \bibfield  {author} {\bibinfo {author} {\bibfnamefont {P.~M.}\ \bibnamefont
  {Paul}}, \bibinfo {author} {\bibfnamefont {E.~S.}\ \bibnamefont {Toma}},
  \bibinfo {author} {\bibfnamefont {P.}~\bibnamefont {Breger}}, \bibinfo
  {author} {\bibfnamefont {G.}~\bibnamefont {Mullot}}, \bibinfo {author}
  {\bibfnamefont {F.}~\bibnamefont {Auge}}, \bibinfo {author} {\bibfnamefont
  {P.}~\bibnamefont {Balcou}}, \bibinfo {author} {\bibfnamefont {H.~G.}\
  \bibnamefont {Muller}},\ and\ \bibinfo {author} {\bibfnamefont
  {P.}~\bibnamefont {Agostini}},\ }\bibfield  {title} {\bibinfo {title}
  {Observation of a {{Train}} of {{Attosecond Pulses}} from {{High Harmonic
  Generation}}},\ }\href {https://doi.org/10.1126/science.1059413} {\bibfield
  {journal} {\bibinfo  {journal} {Science}\ }\textbf {\bibinfo {volume}
  {292}},\ \bibinfo {pages} {1689} (\bibinfo {year} {2001})}\BibitemShut
  {NoStop}%
\bibitem [{\citenamefont {Cho}\ \emph {et~al.}(2019)\citenamefont {Cho},
  \citenamefont {Hwang}, \citenamefont {Nam}, \citenamefont {Bionta},
  \citenamefont {Lassonde}, \citenamefont {Schmidt}, \citenamefont {Ibrahim},
  \citenamefont {L{\'e}gar{\'e}},\ and\ \citenamefont
  {Kim}}]{choTemporalCharacterizationFemtosecond2019}%
  \BibitemOpen
  \bibfield  {author} {\bibinfo {author} {\bibfnamefont {W.}~\bibnamefont
  {Cho}}, \bibinfo {author} {\bibfnamefont {S.~I.}\ \bibnamefont {Hwang}},
  \bibinfo {author} {\bibfnamefont {C.~H.}\ \bibnamefont {Nam}}, \bibinfo
  {author} {\bibfnamefont {M.~R.}\ \bibnamefont {Bionta}}, \bibinfo {author}
  {\bibfnamefont {P.}~\bibnamefont {Lassonde}}, \bibinfo {author}
  {\bibfnamefont {B.~E.}\ \bibnamefont {Schmidt}}, \bibinfo {author}
  {\bibfnamefont {H.}~\bibnamefont {Ibrahim}}, \bibinfo {author} {\bibfnamefont
  {F.}~\bibnamefont {L{\'e}gar{\'e}}},\ and\ \bibinfo {author} {\bibfnamefont
  {K.~T.}\ \bibnamefont {Kim}},\ }\bibfield  {title} {\bibinfo {title}
  {Temporal characterization of femtosecond laser pulses using tunneling
  ionization in the {{UV}}, visible, and mid-{{IR}} ranges},\ }\href
  {https://doi.org/10.1038/s41598-019-52237-y} {\bibfield  {journal} {\bibinfo
  {journal} {Scientific Reports}\ }\textbf {\bibinfo {volume} {9}},\ \bibinfo
  {pages} {1} (\bibinfo {year} {2019})}\BibitemShut {NoStop}%
\bibitem [{\citenamefont {Park}\ \emph {et~al.}(2018)\citenamefont {Park},
  \citenamefont {Kim}, \citenamefont {Cho}, \citenamefont {Hwang},
  \citenamefont {Ivanov}, \citenamefont {Nam},\ and\ \citenamefont
  {Kim}}]{parkDirectSamplingLight2018}%
  \BibitemOpen
  \bibfield  {author} {\bibinfo {author} {\bibfnamefont {S.~B.}\ \bibnamefont
  {Park}}, \bibinfo {author} {\bibfnamefont {K.}~\bibnamefont {Kim}}, \bibinfo
  {author} {\bibfnamefont {W.}~\bibnamefont {Cho}}, \bibinfo {author}
  {\bibfnamefont {S.~I.}\ \bibnamefont {Hwang}}, \bibinfo {author}
  {\bibfnamefont {I.}~\bibnamefont {Ivanov}}, \bibinfo {author} {\bibfnamefont
  {C.~H.}\ \bibnamefont {Nam}},\ and\ \bibinfo {author} {\bibfnamefont {K.~T.}\
  \bibnamefont {Kim}},\ }\bibfield  {title} {\bibinfo {title} {Direct sampling
  of a light wave in air},\ }\href {https://doi.org/10.1364/OPTICA.5.000402}
  {\bibfield  {journal} {\bibinfo  {journal} {Optica}\ }\textbf {\bibinfo
  {volume} {5}},\ \bibinfo {pages} {402} (\bibinfo {year} {2018})}\BibitemShut
  {NoStop}%
\bibitem [{\citenamefont {Bl{\"o}chl}\ \emph {et~al.}(2022)\citenamefont
  {Bl{\"o}chl}, \citenamefont {Sch{\"o}tz}, \citenamefont {Maliakkal},
  \citenamefont {{\v S}reibere}, \citenamefont {Wang}, \citenamefont
  {Rosenberger}, \citenamefont {Hommelhoff}, \citenamefont {Staudte},
  \citenamefont {Corkum}, \citenamefont {Bergues},\ and\ \citenamefont
  {Kling}}]{blochlSpatiotemporalSamplingNearpetahertz2022}%
  \BibitemOpen
  \bibfield  {author} {\bibinfo {author} {\bibfnamefont {J.}~\bibnamefont
  {Bl{\"o}chl}}, \bibinfo {author} {\bibfnamefont {J.}~\bibnamefont
  {Sch{\"o}tz}}, \bibinfo {author} {\bibfnamefont {A.}~\bibnamefont
  {Maliakkal}}, \bibinfo {author} {\bibfnamefont {N.}~\bibnamefont {{\v
  S}reibere}}, \bibinfo {author} {\bibfnamefont {Z.}~\bibnamefont {Wang}},
  \bibinfo {author} {\bibfnamefont {P.}~\bibnamefont {Rosenberger}}, \bibinfo
  {author} {\bibfnamefont {P.}~\bibnamefont {Hommelhoff}}, \bibinfo {author}
  {\bibfnamefont {A.}~\bibnamefont {Staudte}}, \bibinfo {author} {\bibfnamefont
  {P.~B.}\ \bibnamefont {Corkum}}, \bibinfo {author} {\bibfnamefont
  {B.}~\bibnamefont {Bergues}},\ and\ \bibinfo {author} {\bibfnamefont {M.~F.}\
  \bibnamefont {Kling}},\ }\bibfield  {title} {\bibinfo {title} {Spatiotemporal
  sampling of near-petahertz vortex fields},\ }\href
  {https://doi.org/10.1364/OPTICA.459612} {\bibfield  {journal} {\bibinfo
  {journal} {Optica}\ }\textbf {\bibinfo {volume} {9}},\ \bibinfo {pages} {755}
  (\bibinfo {year} {2022})}\BibitemShut {NoStop}%
\bibitem [{\citenamefont {Runge}\ and\ \citenamefont
  {Gross}(1984)}]{rungeDensityFunctionalTheoryTimeDependent1984}%
  \BibitemOpen
  \bibfield  {author} {\bibinfo {author} {\bibfnamefont {E.}~\bibnamefont
  {Runge}}\ and\ \bibinfo {author} {\bibfnamefont {E.~K.~U.}\ \bibnamefont
  {Gross}},\ }\bibfield  {title} {\bibinfo {title} {Density-{{Functional
  Theory}} for {{Time-Dependent Systems}}},\ }\href
  {https://doi.org/10.1103/PhysRevLett.52.997} {\bibfield  {journal} {\bibinfo
  {journal} {Physical Review Letters}\ }\textbf {\bibinfo {volume} {52}},\
  \bibinfo {pages} {997} (\bibinfo {year} {1984})}\BibitemShut {NoStop}%
\bibitem [{\citenamefont {Jensen}\ and\ \citenamefont
  {Madsen}(2021)}]{jensenEdgestateBulklikeLaserinduced2021}%
  \BibitemOpen
  \bibfield  {author} {\bibinfo {author} {\bibfnamefont {S.~V.~B.}\
  \bibnamefont {Jensen}}\ and\ \bibinfo {author} {\bibfnamefont {L.~B.}\
  \bibnamefont {Madsen}},\ }\bibfield  {title} {\bibinfo {title} {Edge-state
  and bulklike laser-induced correlation effects in high-harmonic generation
  from a linear chain},\ }\href {https://doi.org/10.1103/PhysRevB.104.054309}
  {\bibfield  {journal} {\bibinfo  {journal} {Physical Review B}\ }\textbf
  {\bibinfo {volume} {104}},\ \bibinfo {pages} {054309} (\bibinfo {year}
  {2021})}\BibitemShut {NoStop}%
\bibitem [{\citenamefont {{Tancogne-Dejean}}\ and\ \citenamefont
  {Rubio}(2020)}]{tancogne-dejeanParameterfreeHybridlikeFunctional2020}%
  \BibitemOpen
  \bibfield  {author} {\bibinfo {author} {\bibfnamefont {N.}~\bibnamefont
  {{Tancogne-Dejean}}}\ and\ \bibinfo {author} {\bibfnamefont {A.}~\bibnamefont
  {Rubio}},\ }\bibfield  {title} {\bibinfo {title} {Parameter-free hybridlike
  functional based on an extended {{Hubbard}} model: {{DFT}}+{{U}}+{{V}}},\
  }\href {https://doi.org/10.1103/PhysRevB.102.155117} {\bibfield  {journal}
  {\bibinfo  {journal} {Physical Review B}\ }\textbf {\bibinfo {volume}
  {102}},\ \bibinfo {pages} {155117} (\bibinfo {year} {2020})}\BibitemShut
  {NoStop}%
\bibitem [{\citenamefont {{Tancogne-Dejean}}\ \emph
  {et~al.}(2017{\natexlab{a}})\citenamefont {{Tancogne-Dejean}}, \citenamefont
  {M{\"u}cke}, \citenamefont {K{\"a}rtner},\ and\ \citenamefont
  {Rubio}}]{tancogne-dejeanEllipticityDependenceHighharmonic2017}%
  \BibitemOpen
  \bibfield  {author} {\bibinfo {author} {\bibfnamefont {N.}~\bibnamefont
  {{Tancogne-Dejean}}}, \bibinfo {author} {\bibfnamefont {O.~D.}\ \bibnamefont
  {M{\"u}cke}}, \bibinfo {author} {\bibfnamefont {F.~X.}\ \bibnamefont
  {K{\"a}rtner}},\ and\ \bibinfo {author} {\bibfnamefont {A.}~\bibnamefont
  {Rubio}},\ }\bibfield  {title} {\bibinfo {title} {Ellipticity dependence of
  high-harmonic generation in solids originating from coupled intraband and
  interband dynamics},\ }\href {https://doi.org/10.1038/s41467-017-00764-5}
  {\bibfield  {journal} {\bibinfo  {journal} {Nature Communications}\ }\textbf
  {\bibinfo {volume} {8}},\ \bibinfo {pages} {745} (\bibinfo {year}
  {2017}{\natexlab{a}})}\BibitemShut {NoStop}%
\bibitem [{\citenamefont {{Tancogne-Dejean}}\ \emph
  {et~al.}(2017{\natexlab{b}})\citenamefont {{Tancogne-Dejean}}, \citenamefont
  {M{\"u}cke}, \citenamefont {K{\"a}rtner},\ and\ \citenamefont
  {Rubio}}]{tancogne-dejeanImpactElectronicBand2017}%
  \BibitemOpen
  \bibfield  {author} {\bibinfo {author} {\bibfnamefont {N.}~\bibnamefont
  {{Tancogne-Dejean}}}, \bibinfo {author} {\bibfnamefont {O.~D.}\ \bibnamefont
  {M{\"u}cke}}, \bibinfo {author} {\bibfnamefont {F.~X.}\ \bibnamefont
  {K{\"a}rtner}},\ and\ \bibinfo {author} {\bibfnamefont {A.}~\bibnamefont
  {Rubio}},\ }\bibfield  {title} {\bibinfo {title} {Impact of the {{Electronic
  Band Structure}} in {{High-Harmonic Generation Spectra}} of {{Solids}}},\
  }\href {https://doi.org/10.1103/PhysRevLett.118.087403} {\bibfield  {journal}
  {\bibinfo  {journal} {Physical Review Letters}\ }\textbf {\bibinfo {volume}
  {118}},\ \bibinfo {pages} {087403} (\bibinfo {year}
  {2017}{\natexlab{b}})}\BibitemShut {NoStop}%
\bibitem [{\citenamefont {Yu}\ \emph {et~al.}(2019)\citenamefont {Yu},
  \citenamefont {Hansen},\ and\ \citenamefont
  {Madsen}}]{yuEnhancedHighorderHarmonic2019}%
  \BibitemOpen
  \bibfield  {author} {\bibinfo {author} {\bibfnamefont {C.}~\bibnamefont
  {Yu}}, \bibinfo {author} {\bibfnamefont {K.~K.}\ \bibnamefont {Hansen}},\
  and\ \bibinfo {author} {\bibfnamefont {L.~B.}\ \bibnamefont {Madsen}},\
  }\bibfield  {title} {\bibinfo {title} {Enhanced high-order harmonic
  generation in donor-doped band-gap materials},\ }\href
  {https://doi.org/10.1103/PhysRevA.99.013435} {\bibfield  {journal} {\bibinfo
  {journal} {Physical Review A}\ }\textbf {\bibinfo {volume} {99}},\ \bibinfo
  {pages} {013435} (\bibinfo {year} {2019})}\BibitemShut {NoStop}%
\bibitem [{\citenamefont {Bauer}\ and\ \citenamefont
  {Hansen}(2018)}]{bauerHighHarmonicGenerationSolids2018}%
  \BibitemOpen
  \bibfield  {author} {\bibinfo {author} {\bibfnamefont {D.}~\bibnamefont
  {Bauer}}\ and\ \bibinfo {author} {\bibfnamefont {K.~K.}\ \bibnamefont
  {Hansen}},\ }\bibfield  {title} {\bibinfo {title} {High-{{Harmonic
  Generation}} in {{Solids}} with and without {{Topological Edge States}}},\
  }\href {https://doi.org/10.1103/PhysRevLett.120.177401} {\bibfield  {journal}
  {\bibinfo  {journal} {Physical Review Letters}\ }\textbf {\bibinfo {volume}
  {120}},\ \bibinfo {pages} {177401} (\bibinfo {year} {2018})}\BibitemShut
  {NoStop}%
\bibitem [{\citenamefont {Hansen}\ \emph {et~al.}(2018)\citenamefont {Hansen},
  \citenamefont {Bauer},\ and\ \citenamefont
  {Madsen}}]{hansenFinitesystemEffectsHighorder2018}%
  \BibitemOpen
  \bibfield  {author} {\bibinfo {author} {\bibfnamefont {K.~K.}\ \bibnamefont
  {Hansen}}, \bibinfo {author} {\bibfnamefont {D.}~\bibnamefont {Bauer}},\ and\
  \bibinfo {author} {\bibfnamefont {L.~B.}\ \bibnamefont {Madsen}},\ }\bibfield
   {title} {\bibinfo {title} {Finite-system effects on high-order harmonic
  generation: {{From}} atoms to solids},\ }\href
  {https://doi.org/10.1103/PhysRevA.97.043424} {\bibfield  {journal} {\bibinfo
  {journal} {Physical Review A}\ }\textbf {\bibinfo {volume} {97}},\ \bibinfo
  {pages} {043424} (\bibinfo {year} {2018})}\BibitemShut {NoStop}%
\bibitem [{\citenamefont {Jensen}\ \emph {et~al.}(2021)\citenamefont {Jensen},
  \citenamefont {Iravani},\ and\ \citenamefont
  {Madsen}}]{jensenEdgestateinducedCorrelationEffects2021}%
  \BibitemOpen
  \bibfield  {author} {\bibinfo {author} {\bibfnamefont {S.~V.~B.}\
  \bibnamefont {Jensen}}, \bibinfo {author} {\bibfnamefont {H.}~\bibnamefont
  {Iravani}},\ and\ \bibinfo {author} {\bibfnamefont {L.~B.}\ \bibnamefont
  {Madsen}},\ }\bibfield  {title} {\bibinfo {title} {Edge-state-induced
  correlation effects in two-color pump-probe high-order harmonic generation},\
  }\href {https://doi.org/10.1103/PhysRevA.103.053121} {\bibfield  {journal}
  {\bibinfo  {journal} {Physical Review A}\ }\textbf {\bibinfo {volume}
  {103}},\ \bibinfo {pages} {053121} (\bibinfo {year} {2021})}\BibitemShut
  {NoStop}%
\bibitem [{\citenamefont {Ashcroft}\ \emph {et~al.}(1976)\citenamefont
  {Ashcroft}, \citenamefont {W},\ and\ \citenamefont
  {Mermin}}]{ashcroftSolidStatePhysics1976}%
  \BibitemOpen
  \bibfield  {author} {\bibinfo {author} {\bibfnamefont {N.~W.}\ \bibnamefont
  {Ashcroft}}, \bibinfo {author} {\bibfnamefont {A.}~\bibnamefont {W}},\ and\
  \bibinfo {author} {\bibfnamefont {N.~D.}\ \bibnamefont {Mermin}},\
  }\href@noop {} {\emph {\bibinfo {title} {Solid {{State Physics}}}}}\
  (\bibinfo  {publisher} {{Holt, Rinehart and Winston}},\ \bibinfo {year}
  {1976})\BibitemShut {NoStop}%
\bibitem [{\citenamefont {Sundaram}\ and\ \citenamefont
  {Niu}(1999)}]{sundaramWavepacketDynamicsSlowly1999}%
  \BibitemOpen
  \bibfield  {author} {\bibinfo {author} {\bibfnamefont {G.}~\bibnamefont
  {Sundaram}}\ and\ \bibinfo {author} {\bibfnamefont {Q.}~\bibnamefont {Niu}},\
  }\bibfield  {title} {\bibinfo {title} {Wave-packet dynamics in slowly
  perturbed crystals: {{Gradient}} corrections and {{Berry-phase}} effects},\
  }\href {https://doi.org/10.1103/PhysRevB.59.14915} {\bibfield  {journal}
  {\bibinfo  {journal} {Physical Review B}\ }\textbf {\bibinfo {volume} {59}},\
  \bibinfo {pages} {14915} (\bibinfo {year} {1999})}\BibitemShut {NoStop}%
\bibitem [{\citenamefont {Kaneshima}\ \emph {et~al.}(2018)\citenamefont
  {Kaneshima}, \citenamefont {Shinohara}, \citenamefont {Takeuchi},
  \citenamefont {Ishii}, \citenamefont {Imasaka}, \citenamefont {Kaji},
  \citenamefont {Ashihara}, \citenamefont {Ishikawa},\ and\ \citenamefont
  {Itatani}}]{kaneshimaPolarizationResolvedStudyHigh2018}%
  \BibitemOpen
  \bibfield  {author} {\bibinfo {author} {\bibfnamefont {K.}~\bibnamefont
  {Kaneshima}}, \bibinfo {author} {\bibfnamefont {Y.}~\bibnamefont
  {Shinohara}}, \bibinfo {author} {\bibfnamefont {K.}~\bibnamefont {Takeuchi}},
  \bibinfo {author} {\bibfnamefont {N.}~\bibnamefont {Ishii}}, \bibinfo
  {author} {\bibfnamefont {K.}~\bibnamefont {Imasaka}}, \bibinfo {author}
  {\bibfnamefont {T.}~\bibnamefont {Kaji}}, \bibinfo {author} {\bibfnamefont
  {S.}~\bibnamefont {Ashihara}}, \bibinfo {author} {\bibfnamefont {K.~L.}\
  \bibnamefont {Ishikawa}},\ and\ \bibinfo {author} {\bibfnamefont
  {J.}~\bibnamefont {Itatani}},\ }\bibfield  {title} {\bibinfo {title}
  {Polarization-{{Resolved Study}} of {{High Harmonics}} from {{Bulk
  Semiconductors}}},\ }\href {https://doi.org/10.1103/PhysRevLett.120.243903}
  {\bibfield  {journal} {\bibinfo  {journal} {Physical Review Letters}\
  }\textbf {\bibinfo {volume} {120}},\ \bibinfo {pages} {243903} (\bibinfo
  {year} {2018})}\BibitemShut {NoStop}%
\bibitem [{\citenamefont {Liu}\ \emph {et~al.}(2017)\citenamefont {Liu},
  \citenamefont {Li}, \citenamefont {You}, \citenamefont {Ghimire},
  \citenamefont {Heinz},\ and\ \citenamefont
  {Reis}}]{liuHighharmonicGenerationAtomically2017}%
  \BibitemOpen
  \bibfield  {author} {\bibinfo {author} {\bibfnamefont {H.}~\bibnamefont
  {Liu}}, \bibinfo {author} {\bibfnamefont {Y.}~\bibnamefont {Li}}, \bibinfo
  {author} {\bibfnamefont {Y.~S.}\ \bibnamefont {You}}, \bibinfo {author}
  {\bibfnamefont {S.}~\bibnamefont {Ghimire}}, \bibinfo {author} {\bibfnamefont
  {T.~F.}\ \bibnamefont {Heinz}},\ and\ \bibinfo {author} {\bibfnamefont
  {D.~A.}\ \bibnamefont {Reis}},\ }\bibfield  {title} {\bibinfo {title}
  {High-harmonic generation from an atomically thin semiconductor},\ }\href
  {https://doi.org/10.1038/nphys3946} {\bibfield  {journal} {\bibinfo
  {journal} {Nature Physics}\ }\textbf {\bibinfo {volume} {13}},\ \bibinfo
  {pages} {262} (\bibinfo {year} {2017})}\BibitemShut {NoStop}%
\bibitem [{\citenamefont {Jensen}\ and\ \citenamefont
  {Madsen}(2022)}]{jensenPropagationTimeNondipole2022}%
  \BibitemOpen
  \bibfield  {author} {\bibinfo {author} {\bibfnamefont {S.~V.~B.}\
  \bibnamefont {Jensen}}\ and\ \bibinfo {author} {\bibfnamefont {L.~B.}\
  \bibnamefont {Madsen}},\ }\bibfield  {title} {\bibinfo {title} {Propagation
  time and nondipole contributions to intraband high-order harmonic
  generation},\ }\href {https://doi.org/10.1103/PhysRevA.105.L021101}
  {\bibfield  {journal} {\bibinfo  {journal} {Physical Review A}\ }\textbf
  {\bibinfo {volume} {105}},\ \bibinfo {pages} {L021101} (\bibinfo {year}
  {2022})}\BibitemShut {NoStop}%
\bibitem [{\citenamefont {Keiber}\ \emph {et~al.}(2016)\citenamefont {Keiber},
  \citenamefont {Sederberg}, \citenamefont {Schwarz}, \citenamefont
  {Trubetskov}, \citenamefont {Pervak}, \citenamefont {Krausz},\ and\
  \citenamefont {Karpowicz}}]{keiberElectroopticSamplingNearinfrared2016}%
  \BibitemOpen
  \bibfield  {author} {\bibinfo {author} {\bibfnamefont {S.}~\bibnamefont
  {Keiber}}, \bibinfo {author} {\bibfnamefont {S.}~\bibnamefont {Sederberg}},
  \bibinfo {author} {\bibfnamefont {A.}~\bibnamefont {Schwarz}}, \bibinfo
  {author} {\bibfnamefont {M.}~\bibnamefont {Trubetskov}}, \bibinfo {author}
  {\bibfnamefont {V.}~\bibnamefont {Pervak}}, \bibinfo {author} {\bibfnamefont
  {F.}~\bibnamefont {Krausz}},\ and\ \bibinfo {author} {\bibfnamefont
  {N.}~\bibnamefont {Karpowicz}},\ }\bibfield  {title} {\bibinfo {title}
  {Electro-optic sampling of near-infrared waveforms},\ }\href
  {https://doi.org/10.1038/nphoton.2015.269} {\bibfield  {journal} {\bibinfo
  {journal} {Nature Photonics}\ }\textbf {\bibinfo {volume} {10}},\ \bibinfo
  {pages} {159} (\bibinfo {year} {2016})}\BibitemShut {NoStop}%
\bibitem [{\citenamefont {Putnam}\ \emph {et~al.}(2017)\citenamefont {Putnam},
  \citenamefont {Hobbs}, \citenamefont {Keathley}, \citenamefont {Berggren},\
  and\ \citenamefont
  {K{\"a}rtner}}]{putnamOpticalfieldcontrolledPhotoemissionPlasmonic2017}%
  \BibitemOpen
  \bibfield  {author} {\bibinfo {author} {\bibfnamefont {W.~P.}\ \bibnamefont
  {Putnam}}, \bibinfo {author} {\bibfnamefont {R.~G.}\ \bibnamefont {Hobbs}},
  \bibinfo {author} {\bibfnamefont {P.~D.}\ \bibnamefont {Keathley}}, \bibinfo
  {author} {\bibfnamefont {K.~K.}\ \bibnamefont {Berggren}},\ and\ \bibinfo
  {author} {\bibfnamefont {F.~X.}\ \bibnamefont {K{\"a}rtner}},\ }\bibfield
  {title} {\bibinfo {title} {Optical-field-controlled photoemission from
  plasmonic nanoparticles},\ }\href {https://doi.org/10.1038/nphys3978}
  {\bibfield  {journal} {\bibinfo  {journal} {Nature Physics}\ }\textbf
  {\bibinfo {volume} {13}},\ \bibinfo {pages} {335} (\bibinfo {year}
  {2017})}\BibitemShut {NoStop}%
\bibitem [{\citenamefont {Keathley}\ \emph {et~al.}(2019)\citenamefont
  {Keathley}, \citenamefont {Putnam}, \citenamefont {Vasireddy}, \citenamefont
  {Hobbs}, \citenamefont {Yang}, \citenamefont {Berggren},\ and\ \citenamefont
  {K{\"a}rtner}}]{keathleyVanishingCarrierenvelopephasesensitiveResponse2019}%
  \BibitemOpen
  \bibfield  {author} {\bibinfo {author} {\bibfnamefont {P.~D.}\ \bibnamefont
  {Keathley}}, \bibinfo {author} {\bibfnamefont {W.~P.}\ \bibnamefont
  {Putnam}}, \bibinfo {author} {\bibfnamefont {P.}~\bibnamefont {Vasireddy}},
  \bibinfo {author} {\bibfnamefont {R.~G.}\ \bibnamefont {Hobbs}}, \bibinfo
  {author} {\bibfnamefont {Y.}~\bibnamefont {Yang}}, \bibinfo {author}
  {\bibfnamefont {K.~K.}\ \bibnamefont {Berggren}},\ and\ \bibinfo {author}
  {\bibfnamefont {F.~X.}\ \bibnamefont {K{\"a}rtner}},\ }\bibfield  {title}
  {\bibinfo {title} {Vanishing carrier-envelope-phase-sensitive response in
  optical-field photoemission from plasmonic nanoantennas},\ }\href
  {https://doi.org/10.1038/s41567-019-0613-6} {\bibfield  {journal} {\bibinfo
  {journal} {Nature Physics}\ ,\ \bibinfo {pages} {1}} (\bibinfo {year}
  {2019})}\BibitemShut {NoStop}%
\bibitem [{\citenamefont {Rybka}\ \emph {et~al.}(2016)\citenamefont {Rybka},
  \citenamefont {Ludwig}, \citenamefont {Schmalz}, \citenamefont {Knittel},
  \citenamefont {Brida},\ and\ \citenamefont
  {Leitenstorfer}}]{rybkaSubcycleOpticalPhase2016}%
  \BibitemOpen
  \bibfield  {author} {\bibinfo {author} {\bibfnamefont {T.}~\bibnamefont
  {Rybka}}, \bibinfo {author} {\bibfnamefont {M.}~\bibnamefont {Ludwig}},
  \bibinfo {author} {\bibfnamefont {M.~F.}\ \bibnamefont {Schmalz}}, \bibinfo
  {author} {\bibfnamefont {V.}~\bibnamefont {Knittel}}, \bibinfo {author}
  {\bibfnamefont {D.}~\bibnamefont {Brida}},\ and\ \bibinfo {author}
  {\bibfnamefont {A.}~\bibnamefont {Leitenstorfer}},\ }\bibfield  {title}
  {\bibinfo {title} {Sub-cycle optical phase control of nanotunnelling in the
  single-electron regime},\ }\href {https://doi.org/10.1038/nphoton.2016.174}
  {\bibfield  {journal} {\bibinfo  {journal} {Nature Photonics}\ }\textbf
  {\bibinfo {volume} {10}},\ \bibinfo {pages} {667} (\bibinfo {year}
  {2016})}\BibitemShut {NoStop}%
\bibitem [{\citenamefont {Schoetz}\ \emph {et~al.}(2019)\citenamefont
  {Schoetz}, \citenamefont {Wang}, \citenamefont {Pisanty}, \citenamefont
  {Lewenstein}, \citenamefont {Kling},\ and\ \citenamefont
  {Ciappina}}]{schoetzPerspectivePetahertzElectronics2019}%
  \BibitemOpen
  \bibfield  {author} {\bibinfo {author} {\bibfnamefont {J.}~\bibnamefont
  {Schoetz}}, \bibinfo {author} {\bibfnamefont {Z.}~\bibnamefont {Wang}},
  \bibinfo {author} {\bibfnamefont {E.}~\bibnamefont {Pisanty}}, \bibinfo
  {author} {\bibfnamefont {M.}~\bibnamefont {Lewenstein}}, \bibinfo {author}
  {\bibfnamefont {M.~F.}\ \bibnamefont {Kling}},\ and\ \bibinfo {author}
  {\bibfnamefont {M.~F.}\ \bibnamefont {Ciappina}},\ }\bibfield  {title}
  {\bibinfo {title} {Perspective on {{Petahertz Electronics}} and {{Attosecond
  Nanoscopy}}},\ }\href {https://doi.org/10.1021/acsphotonics.9b01188}
  {\bibfield  {journal} {\bibinfo  {journal} {ACS Photonics}\ }\textbf
  {\bibinfo {volume} {6}},\ \bibinfo {pages} {3057} (\bibinfo {year}
  {2019})}\BibitemShut {NoStop}%
\bibitem [{\citenamefont
  {Bauer}(2017)}]{bauerComputationalStrongfieldQuantum2017}%
  \BibitemOpen
  \bibinfo {editor} {\bibfnamefont {D.}~\bibnamefont {Bauer}},\ ed.,\
  \href@noop {} {\emph {\bibinfo {title} {Computational Strong-Field Quantum
  Dynamics: Intense Light-Matter Interactions}}},\ De {{Gruyter}} Graduate\
  (\bibinfo  {publisher} {{Walter de Gruyter GmbH}},\ \bibinfo {address}
  {{Berlin ; Boston}},\ \bibinfo {year} {2017})\BibitemShut {NoStop}%
\bibitem [{\citenamefont {Kosloff}\ and\ \citenamefont
  {Kosloff}(1986)}]{kosloffAbsorbingBoundariesWave1986}%
  \BibitemOpen
  \bibfield  {author} {\bibinfo {author} {\bibfnamefont {R.}~\bibnamefont
  {Kosloff}}\ and\ \bibinfo {author} {\bibfnamefont {D.}~\bibnamefont
  {Kosloff}},\ }\bibfield  {title} {\bibinfo {title} {Absorbing boundaries for
  wave propagation problems},\ }\href
  {https://doi.org/10.1016/0021-9991(86)90199-3} {\bibfield  {journal}
  {\bibinfo  {journal} {Journal of Computational Physics}\ }\textbf {\bibinfo
  {volume} {63}},\ \bibinfo {pages} {363} (\bibinfo {year} {1986})}\BibitemShut
  {NoStop}%
\bibitem [{\citenamefont {Kilen}\ \emph {et~al.}(2020)\citenamefont {Kilen},
  \citenamefont {Kolesik}, \citenamefont {Hader}, \citenamefont {Moloney},
  \citenamefont {Huttner}, \citenamefont {Hagen},\ and\ \citenamefont
  {Koch}}]{kilenPropagationInducedDephasing2020}%
  \BibitemOpen
  \bibfield  {author} {\bibinfo {author} {\bibfnamefont {I.}~\bibnamefont
  {Kilen}}, \bibinfo {author} {\bibfnamefont {M.}~\bibnamefont {Kolesik}},
  \bibinfo {author} {\bibfnamefont {J.}~\bibnamefont {Hader}}, \bibinfo
  {author} {\bibfnamefont {J.~V.}\ \bibnamefont {Moloney}}, \bibinfo {author}
  {\bibfnamefont {U.}~\bibnamefont {Huttner}}, \bibinfo {author} {\bibfnamefont
  {M.~K.}\ \bibnamefont {Hagen}},\ and\ \bibinfo {author} {\bibfnamefont
  {S.~W.}\ \bibnamefont {Koch}},\ }\bibfield  {title} {\bibinfo {title}
  {Propagation {{Induced Dephasing}} in {{Semiconductor High-Harmonic
  Generation}}},\ }\href {https://doi.org/10.1103/PhysRevLett.125.083901}
  {\bibfield  {journal} {\bibinfo  {journal} {Physical Review Letters}\
  }\textbf {\bibinfo {volume} {125}},\ \bibinfo {pages} {083901} (\bibinfo
  {year} {2020})}\BibitemShut {NoStop}%
\bibitem [{\citenamefont {Yamada}\ and\ \citenamefont
  {Yabana}(2021)}]{yamadaDeterminingOptimumThickness2021}%
  \BibitemOpen
  \bibfield  {author} {\bibinfo {author} {\bibfnamefont {S.}~\bibnamefont
  {Yamada}}\ and\ \bibinfo {author} {\bibfnamefont {K.}~\bibnamefont
  {Yabana}},\ }\bibfield  {title} {\bibinfo {title} {Determining the optimum
  thickness for high harmonic generation from nanoscale thin films: {{An}} ab
  initio computational study},\ }\href
  {https://doi.org/10.1103/PhysRevB.103.155426} {\bibfield  {journal} {\bibinfo
   {journal} {Physical Review B}\ }\textbf {\bibinfo {volume} {103}},\ \bibinfo
  {pages} {155426} (\bibinfo {year} {2021})}\BibitemShut {NoStop}%
\end{thebibliography}%



\end{document}